\documentclass[pdflatex,sn-mathphys-num]{sn-jnl}% Math and Physical Sciences Numbered Reference Style
\usepackage{graphicx}%
\usepackage{multirow}%
\usepackage{amsmath,amssymb,amsfonts}%
\usepackage{amsthm}%
\usepackage{mathrsfs}%
\usepackage[title]{appendix}%
\usepackage{xcolor}%
\usepackage{textcomp}%
\usepackage{manyfoot}%
\usepackage{booktabs}%
\usepackage{algorithm}%
\usepackage{algorithmicx}%
\usepackage{algpseudocode}%
\usepackage{listings}%
\theoremstyle{thmstyleone}%
\theoremstyle{thmstyletwo}%

\theoremstyle{thmstylethree}%

\begin{document}

\title[Article Title]{Vortex rings as hydrodynamic dispersants for mitigating underwater oil spills}


\author[]{\fnm{Siddhant} \sur{Jain}}

\author[]{\fnm{Saini Jatin} \sur{Rao}}
%\equalcont{}

\author[*]{\fnm{Saptarshi} \sur{Basu}}\email{sbasu@iisc.ac.in}
%\equalcont{}

\affil[]{\orgdiv{Dept. of Mechanical Engineering}, \orgname{Indian Institute of Science}, \orgaddress{\city{Bengaluru}, \postcode{560012}, \country{India}}}


\abstract{The usage of chemical dispersant to promote the breakup of oil droplets remains one of the crucial response strategy towards controlling the rapidly rising droplets during an underwater oil spillage disaster like the \textit{Deepwater horizon}. However, the detrimental effects of dispersant on the marine life are now well recognized. Here, we demonstrate that vortex ring presents an environmental friendly alternative to the debatable usage of chemical dispersant that primarily aids droplet break-up. We propose two mitigating strategies through the application of vortex rings: (i) atomization of large oil droplets into smaller droplets, which enhances natural dispersion and biodegradation (ii) targeted relocation of spilled oil droplets from ecologically sensitive regions to safer remote locations during minor spill events. This proof-of-concept study demonstrates that vortex rings can efficiently atomize underwater oil droplets while simultaneously transporting them over long distances. The efficiency strongly depends on the droplet size distribution and the strength of the vortex ring. Using time scale analysis and force balances arguments, we propose droplet response and atomization criteria by a traversing vortex ring. Beyond atomization and droplet transport which are center to many engineering processes, the work finds high relevance in the area of bubble turbulence, agricultural industry, multiphase reactors, combustion process, atmospheric fluid dynamics and oil extraction from rocks, to name a few.}




\maketitle

\section{Introduction}\label{sec1}

The \textit{Deepwater Horizon} (DwH) spill is recognized as one of the most colossal spillage event where plume of oil jetted towards the sea surface from a depth of $\sim$ 1500 m \citep{Lubchenco2012, Passow2021} triggering a plethora of chemical and physical events including biodegradation, dissolution, weathering, interaction with organic matter and outer environment, recoalescence, atomization, transport and dispersion \citep{Boufadel2023}.The consequences of such large-scale spills have been profoundly detrimental, posing threats to marine ecosystems and potentially impacting human health \cite{Fisher2014, Snyder2015,Laffon2016, Echols2016, Grosell2021}.

One of the primary responses to the Deepwater Horizon (DwH) spill was the application of the chemical dispersant Corexit at both the blowout site and the sea surface \citep{Gros2017, Passow2021, Boufadel2023}. Dispersants are amphiphilic compounds that reduce the oil water interfacial tension \citep{Balaji_PRL, Cressey2010}, promoting the break-up of oil into smaller droplets under ocean currents, turbulence, and mixing. Smaller droplets possess a higher surface area-to-volume ratio (SA:V) and lower buoyancy, leading to enhanced dissolution, prolonged residence time in the water column, and improved microbial biodegradation due to greater hydrocarbon bioavailability and larger colonization area \citep{Kleindienst2015, Passow2021, Boufadel2023, Lee2013, McFarlin2014}. Simulations further showed that sub-sea dispersant injection (SSDI) reduced benzene and other soluble volatile organic compound emissions by nearly 2000 fold into the atmosphere \citep{Gros2017,Zhao2021}. In addition, smaller droplets of oil reaching the surface forms thinner slicks that are easier to manage and reduce shoreline contamination compared with thicker slicks \citep{JOHANSEN2013}.  

Nevertheless, mitigation strategies are contextual to each spill \citep{Passow2021}. For example, in the case of DwH, the spillage was huge and concerns about safety of surface worker, beaches of Mexico and marshes led to the usage of dispersant at the surface as well as at depth \citep{John2016, Zhao2021}. It was preferred to let a large portion of the oil stay afloat underwater rather than rising to the surface. In such scenario, droplets of smaller sizes are preferred over larger ones for reasons discussed above. But the smaller droplets which are a by-product of the usage of chemical dispersant enhance the potential to cause detrimental effects on the marine life. Exposure to Corexit 9500A, either alone or in combination with crude oil, significantly reduces the survival and settlement success of coral larvae, posing a threat to coral recruitment and reef recovery \citep{Goodbody-Gringley2013}. In European sea bass (\textit{Dicentrarchus labrax}), chemically dispersed oil has been shown to cause persistent impairment of hypoxia tolerance, even after exposure has ceased, indicating long-term physiological stress \citep{Zhang2017}. In addition, juvenile European sea bass exhibit altered anti-predator behaviour and disrupted metabolic responses following exposure to dispersant-treated oil \citep{Aimon2022}. Toxicity assessments using moon jellyfish (\textit{Aurelia aurita}) ephyrae demonstrated that dispersed Macondo (DwH) oil induced pronounced sublethal effects, including reduced pulsation and impaired survivorship among other physical abnormalities \citep{Echols2016}. Furthermore, exposure to Corexit 9500A caused structural damage to the gills of the blue crab (\textit{Callinectes sapidus}) and impaired ion transport function, compromising osmoregulatory capacity essential for survival \citep{Weiner2021}. Collectively, these studies and more \citep{Wise2011-dy, almeda2014ingestion} demonstrate that while dispersants enhance oil dispersion, they also impose substantial ecological and physiological risks across diverse marine taxa. Hence, it is crucial to find alternative solutions for creating droplet dispersions during spillage.

\begin{figure}[H]
\centering
\includegraphics[width=1\textwidth]{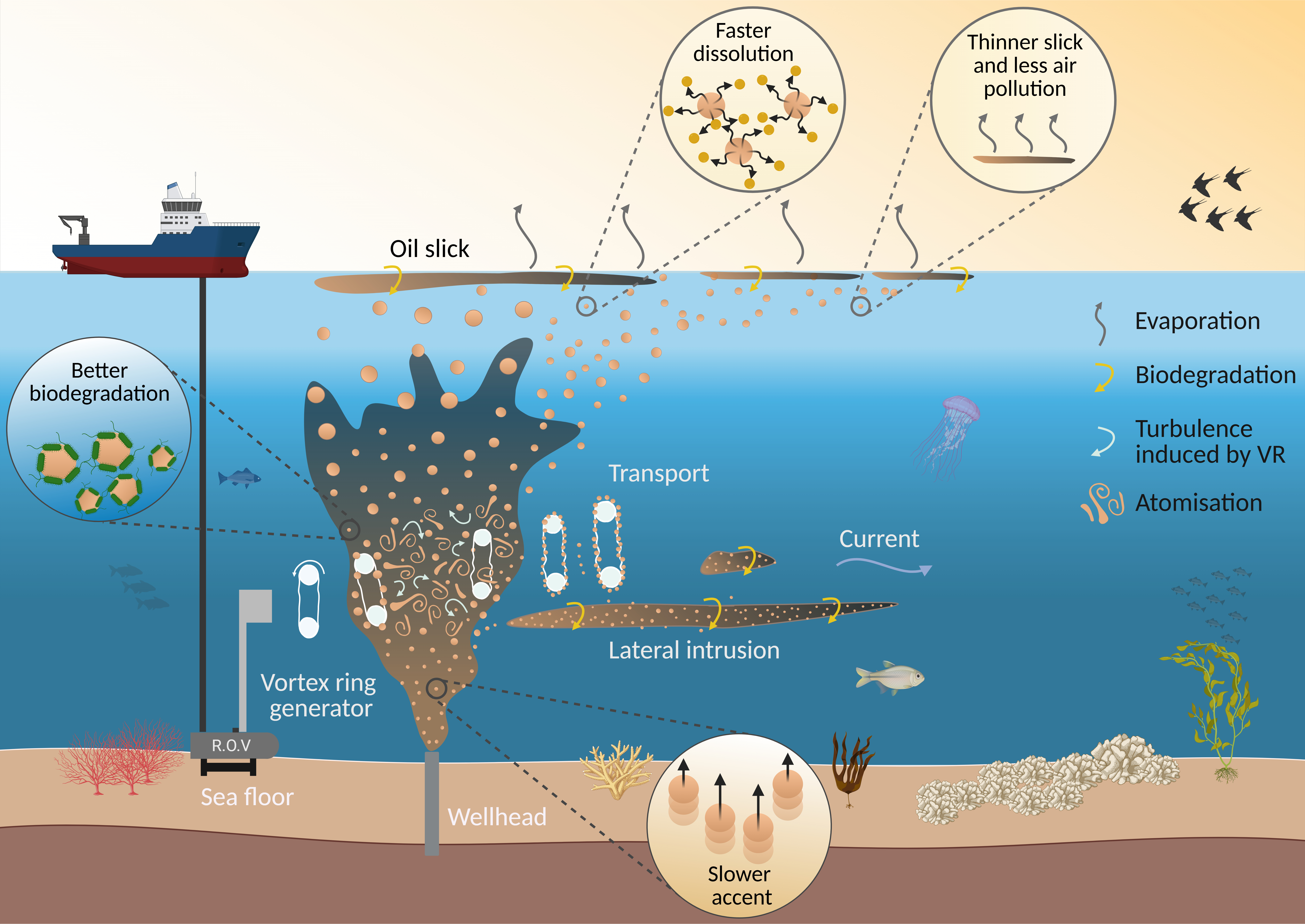}
\caption{A schematic representation of an underwater oil spillage event and the proposed idea to break larger droplets into smaller ones through vortex rings along with their subsequent transport. Oil plume containing droplets can be seen to rising from the blowout at the wellhead that ultimately forms oil slicks at the sea surface. A lateral intrusion layer is shown along the direction of the current. R.O.V. refers to remotely operated vehicle. The advantages of smaller droplets over larger droplets are depicted in the extruding callouts. The figure is created using Biorender. }\label{fig1}
\end{figure}

Instead of using a debatable chemical process \citep{Cressey2010, Kleindienst2015} to obtain smaller droplets, we propose the usage of localized vortex ring of high Reynolds number ($Re_{\Gamma} = \Gamma/\nu$, where $\Gamma$ is the average circulation measured at the interaction site and $\nu$ is the kinematic viscosity of water) for atomization and subsequent transport of underwater spilled droplets. Vortex rings are simpler and stable form of rotating fluid that self propels once ejected impulsively from a tube \citep{SHARIFF_1998, Jain2023}. It has a translating as well as rotating component to it making it an apt bet for mixing and transport problems. In the case of a multiphase scenario, if the inertial forcing (here, applied through vortex rings) perturbs the interface with sufficient energy, it leads to atomization at the interface through various modes. With this background, in the present work, a horizontally traversing vortex ring is made to interact with a swarm of oil droplets for three different droplet distributions: Distribution-I (D-I), Distribution-II (D-II),Distribution-III (D-III) where D-I, D-II and D-III predominantly contains coarse, medium and fine sized-droplets respectively. The configuration is motivated by realistic underwater spillage and two distinct but related capabilities of the vortex ring. Firstly, the rotating vortex core and the translational structure imposes axial and azimuthal stretching on entrained droplets, driving their atomization through a purely non-chemical, physical process. And secondly, the convective capability of vortex ring offers a mechanism for the targeted relocation of oil droplets, particularly in minor sub-sea spill scenarios such as small leaks near coral reefs  where droplets must be moved away from ecologically sensitive regions toward safer sites. The former presents a compelling alternative to the controversial use of chemical dispersants such as Corexit during oil spill response. The underlying physics, being general to any dispersed multiphase system, extends naturally to gas bubbles and other particle/bubble laden flows.

Transport of dispersed phases like bubbles, droplets, and particulates is central to a striking range of natural and engineered processes, spanning drug and pesticide delivery \citep{Herpin2017, mouallem2021,Li2025}, species mixing \citep{Sudarsan2006}, particle-laden turbulence \citep{Eaton1994, Coletti2022}, nanofluid-based heat transfer \citep{Xuan2000}, mass transfer \citep{DiGiorgio2026}, gas-phase catalysis \citep{Kim2020}, emulsion atomization \citep{Yi2022}, and $CO_2$ exchange at the ocean–atmosphere interface \citep{Rodriguez2025}. Among the diverse mechanisms that drive such transport, coherent vortex rings in particular offer a uniquely tunable and spatially precise mean of control. Vortex ring reconnection has been shown to direct particulate drugs precisely toward targeted channel side walls \citep{mouallem2021}. Interaction with solid inertial particle exhibits three distinct regimes: simple deviation, strong deviation, and capture \citep{deAquino2025}. At larger scales, a spark discharge driven underwater bubble ring has been used to transport ferromagnetic (45 $\mu m$) and glass (800 $\mu m$) particles, generating circulation values as high as 8000 $cm^2/s$ and demonstrating potential of highly energetic bubble rings for efficient particle transport \citep{Wang2025}. Successful transport of particles of lower density through vortex rings has been demonstrated \citep{Watanabe}. Beyond transport applications, bubble vortex interactions are of broad engineering significance, underlying phenomena such as drag reduction in bubbly turbulent flows \citep{Jha_Govardhan_2015} and bubble dynamics in wave breaking \citep{Deane2002}. Even microscopic bubbles captured within a vortex core have been shown to substantially distort the ring's structure and vorticity field \citep{SRIDHAR1999JFM}. More recently, under the turbulent conditions generated by head-on vortex ring collisions, secondary break-up of daughter droplets has been shown to be mediated by sub-bubble-scale eddies, challenging the classical Kolmogorov–Hinze picture of droplet break-up \citep{Qi2022}. Co-axial collision of a relatively large droplet ($d_{ring}/d_{d} \sim 2$, i.e., ratio of ring diameter to the droplet diameter) with vortex ring has been reported \citep{Sharma_Singh_Basu_2021} where it is showed that a droplet is first deformed due to high pressure from vortex followed by its stretching and engulfment of droplet within the ring leading to its atomization.  Numerical simulations have highlighted trapping of $\leq 0.4 mm$ droplets in the counter rotating vortex generated due to cross jet-flow interaction \citep{Daskiran2021}. Vortex rings have previously been demonstrated to remove oil from porous surfaces, with the unique advantage of simultaneously cleaning both sides of the substrate owing to their coupled translational and rotational kinematics \citep{Jain2023}. Although a lot has been done in context of drop laden flows \citep{Rodriguez2025}, targeted interaction of vortex rings with rising droplets has not been explored which presents a significant alternative to the dispersant usage.

\begin{figure}[H]
\centering
\includegraphics[width=1\textwidth]{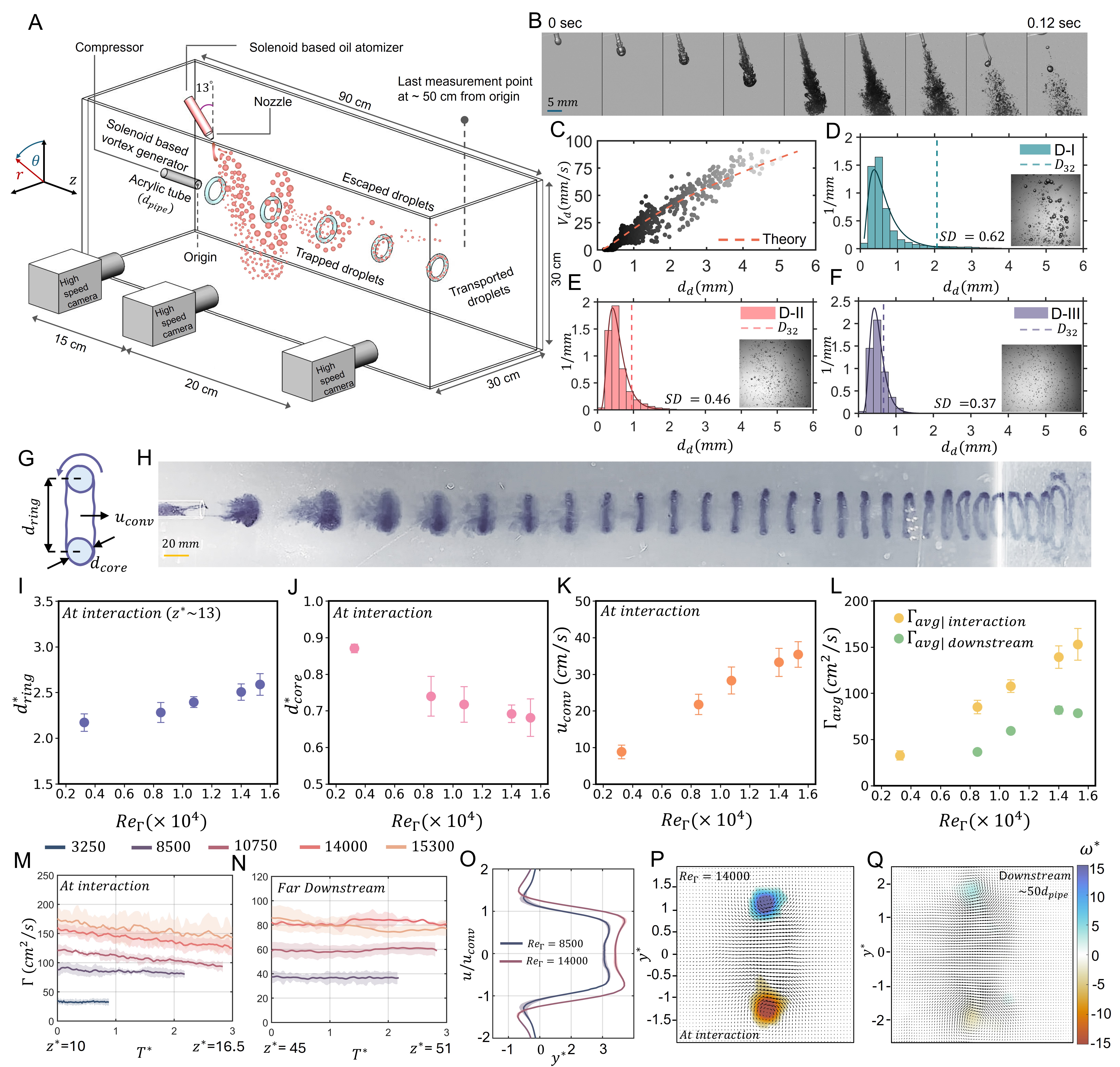}
\caption{(A) A schematic (not to scale) of the experimental setup (B) Time series snapshots of the oil jet mimicking oil spillage (C) Measured rising velocity of droplets and the theoretically estimated trend (D-F) Droplet size distribution for available droplets with (D) $D_{32}$ = 2.06 mm for D-I (E) $D_{32}$ = 0.95 mm for D-II (F) $D_{32}$ = 0.67 mm for D-III and a representative snapshot in the inset (G) Nomenclature of a typical vortex ring (H) Superimposition of high-speed camera images showing the trajectory (dyed with ink) of a vortex ring convecting horizontally. (I-J) The variation of vortex ring diameter ($d_\mathrm{ring}^*$) and core diameter ($d_\mathrm{ring}^*$) at the interaction site $z^*\sim13$ (K) The convective velocity of the vortex ring for different $Re_\Gamma$ (L) The average $\Gamma$ values for each $Re_\Gamma$ at interaction and downstream location ($z^*\sim50$) (M-N) Variation of the instantaneous $\Gamma$ with time at the interaction site and downstream. The spatial distances of the measurements from the origin are mentioned along $x$ axis (O) The velocity profile plotted across the line joining the two cores for $Re_\Gamma = 8500$ and $14000$ (P-Q) The normalised vorticity contours overlaid over the vector field at the interaction site and far downstream location. The length scales are rendered non-dimensional using $d_\mathrm{pipe}$ and time scales using the scale $u_\mathrm{conv}/d_\mathrm{pipe}$.}
\label{fig2}
\end{figure}

The proposed concept along with advantages of having smaller sized droplets is schematically represented in figure~\ref{fig1}. Five vortex ring strengths spanning $Re_\Gamma = 3250 - 15300$ are tested against the three droplet size groups. Olive oil (surface tension with respect to water, $\sigma$ = 17.6 $mN/m$, dynamic viscosity, $\mu$ = 70 $mPas$ and density, $\rho$ = 920 $kg/m^3$) is used for multi-droplet interaction experiments while  additionally JetA fuel ($\sigma$ with water = 6.9 $mN/m$, $\mu$ = 1.22 c and density, $\rho$ = 780 $kg/m^3$) is investigated for single-droplet interactions (shown in supplementary videos) to isolate the role of viscosity and surface tension on break-up dynamics. The droplet population is tracked through four stages marking states before, during and after the interaction: (1) available, (2) affected, (3) trapped, and (4) transported droplets. Sauter mean diameter ($D_{32}=\sum d_i^3/\sum d_i^2$, where $d_i$ represents droplet diameter \citep{kowalczuk2016physical, keshavarz2016}) along with probability distribution function of droplet distribution at each stage are used to characterize the evolution of the population through the whole process and quantify volume weighted transport. The trapping, transport, and net volumetric efficiencies are reported across all cases. Finally, using force balance arguments and timescale analysis, predictive criteria are derived for the droplet response and the threshold diameter for atomization onset as a function of vortex strength.

\section{Results}\label{sec2}
\subsection{Observations on the ring property and spilled oil jet}\label{subsec2.1}

A pressurized chunk of water when ejected through a pipe ($d_{pipe}=10mm$) into a quieter pool of water rolls up due to the separation of the flow at the pipe edge leading to generation of vortex ring that self-propels once formed \citep{SHARIFF_1998, jain2026EXIF}(additional details can be found in Methods section~\ref{sec4}). Now, when the chunk of fluid is beyond optimum condition, the out flowing fluid forms a pulsed liquid jet\citet{SHARIFF_1998}. The latter is used to generated underwater oil-jet in water that mimics a local oil spillage. Figure~\ref{fig2}(A) depicts the experimental setup developed for studying the atomization and transport of spilled oil using vortex rings. Figure~\ref{fig2}(B) shows one of the cases where the pulse of oil jet is ejected out into the water. The jet initially travels against buoyancy before reversing and ascending at the speeds shown in figure~\ref{fig2}(C). Larger droplets travel the shortest distance against buoyancy before reversing and rising. The measured terminal velocities ($v_d$) shows excellent agreement with theoretical estimates (see Supplementary Sheet, Section A for theory). As expected, larger droplets rise faster due to greater buoyancy, naturally segregating the droplets by their sizes. This enables experiments at three distinct time instances, producing three droplet distributions: D-I, D-II, and D-III that the vortex ring encounters as shown in figure~\ref{fig2}(D-F) with decreasing Sauter mean diameter ($D_\mathrm{32}$). The nomenclature for the vortex ring along with an overlaid images of horizontally convecting vortex ring is depicted in figure~\ref{fig2}(G) and (H) respectively. Figure~\ref{fig2}(I-J) depicts the variation of ring diameter ($d_\mathrm{ring}$) and core diameter ($d_{core}$) with the vortex strength. It is observed that $d_\mathrm{ring}$ increases with the $Re_\Gamma$ while the core goes on decreasing in size (also seen in literature \citep{Weigand1997}). Figure~\ref{fig2}(K) shows that increasing trend of convection velocity (calculated using core tracking method, where the core is identified through $\Gamma_2$ method \citep{jain2025Vwall}, refer to section B in supplementary sheet for details) with $Re_\Gamma$. The average circulation, $\Gamma_\mathrm{avg}$ (calculation can be found in Methods) is plotted against the $Re_\Gamma$ for the interaction site and far downstream in figures~\ref{fig2}(L). Further, figures~\ref{fig2}(M) and (N) shows the variation of $\Gamma$ with normalized time ($T^*=tu_\mathrm{conv}/d_\mathrm{pipe}$) at the two locations. Evidently, the circulation decreases more for higher $Re_\Gamma$ because of larger dissipative effects at high $Re_\Gamma$. Far downstream, the $Re_\Gamma$ values significantly dips with more stable variation in time. The velocity profile across the two cores of the vortex ring and the vorticity contour at the interaction site and far downstream are shown in figures~\ref{fig2}(O-Q) respectively. These figure ascertain the symmetric behavior of the vortex ring with minimal sheddin in the wake.

\subsection{The Interaction}\label{subsec2.2}

The interaction of vortex ring with the spilled oil drops produces different outcomes like droplet deviation, deformation, engulfment, transport, atomization, ejection and their combinations (check supplementary videos, SV 1-19, with the video details in Section J of the supplementary sheet). The primary non-dimensional numbers governing these events are Stokes number which is the ratio of droplet response time to the flow time scale ($St = \tau_d/\tau_f=(\rho_d d_d^2/18 \mu_c)/(d_\mathrm{ring}/u_\mathrm{conv})$, where $\tau_d$ is the droplet response time, $\tau_f$ is the flow time scale, $\rho_d$ is the density of the dispersed phase, $d_d$ is the droplet diameter and $\mu_c$ is the dynamics viscosity of the continuous phase i.e. water) and the Weber number which is the ratio of inertial force to the surface tension force ($We = \rho u^2d_d/\sigma,$ with $u = \Gamma/2\pi r_\mathrm{core}$, where $\sigma$ is the surface tension and $r_\mathrm{core}$ is the radius of the core). For $We$, the velocity scale used is the vortex induced velocity resulting in a range of $We$ = 15-145 depending on the droplet size where atomization is observed. For very small droplets, the $We$ values will be extremely small which cannot be ascertained through present measurements.

\begin{figure}[H]
\centering
\includegraphics[width=1\textwidth]{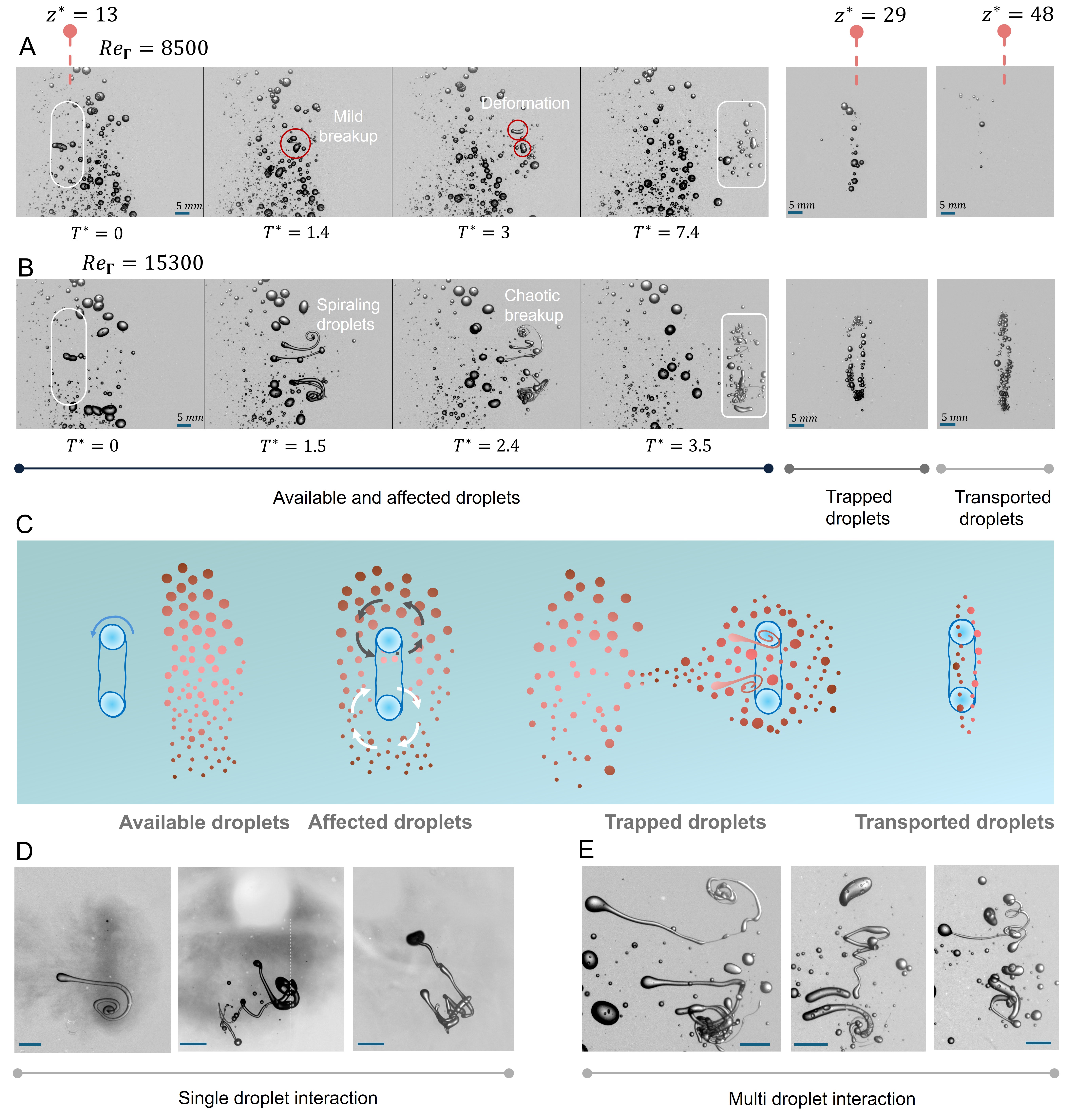}
\caption{Vortex ring with (A) $Re_\Gamma = 8500$ (B) $Re_\Gamma = 15300$ interacting with oil jet having D-I distribution. The interaction with lower $Re_\Gamma$ results in mild break-up of droplets with minimal stretching unlike for higher $Re_\Gamma$  where intense spiraling and break-up are observed. The white zone depicting the vortex ring is not to scale and is meant only for intuitive understanding (C) A two-dimensional schematic depicting various stages of the interaction phenomenon. The white region in the second instance depicts the hollow region created by the vortex as it crosses the cluster of smaller sized droplets. Stretching and spiraling of the droplet along with a trailing wake (as explained in section~\ref{subsec2.2} is also shown in the third instance). Snapshots from (D) single droplet experiments on olive oil captured form side-view (first image) and front view (last two images) and (E) multi droplet experiments showing axial stretching, spiraling and azimuthal stretching of droplets. All the scale bar represents 5mm.}\label{fig3}
\end{figure}

When a vortex ring crosses a swarm of droplets, it influences a much larger volume compared to its own due to the disturbance it induces. Based on this, we define affected droplets as follows: (see section C in supplementary sheet for thresholding details)

\begin{equation}
    d_d = 
\begin{cases}
    \mathrm{affected } & \text{if } \left|\Delta z\right| \geq d_{core} \mid \mathrm{atomized} \\
    \mathrm{unaffected} & \text{if } \left|\Delta z\right| < d_{core} \\
\end{cases}
\label{eqn 1}
\end{equation}
where $\Delta z = z(t)-z_0$ is the displacement of the droplet. This definition is independent of the size of the droplets and dependent on the flow property since, we present most of the data by varying the $Re_\Gamma$. It is important to note that the relative orientation between the droplet and the vortex-ring core can play a crucial role in the interaction and transport process. To reduce its influence on the global statistics, a large number of data sets are processed ($\geq$ 7 runs for each unique case). Figure~\ref{fig3}(A,B) depicts the interaction process for two strengths of vortex ring interacting with D-I size distribution (For video check SV 1-4). As mentioned above, we divide the interaction into four stages depicted in Figure~\ref{fig3}(C). Camera 1 captures the first two stages i.e. available and affected droplets followed by camera 2 and 3 capturing the trapped and transported droplets respectively. 

We observe that vortex with $Re_\Gamma = 8500$ produce mild break-up of sufficiently larger droplet along with deforming them followed by their trapping. Due to smaller $We$ (among all the strengths considered here) which depends on the $\sim\Gamma^2d_\mathrm{d} $, vortex with less strength is not able to overcome the capillary resistance provided by the droplet thereby producing less atomization. In most cases, this vortex being weaker, succumbs to the drag induced by the droplets on it, resulting in ejection of large number of droplets before it reaches far downstream $(z^*\sim 50)$ that is the target set for the present experiments (can be seen from transport efficiency plot in section~\ref{subsec2.4}). In contrast, vortex rings with larger $Re_\Gamma$ generate sufficiently high inertial stresses to overcome capillary resistance, resulting in intense primary atomization through droplet stretching and spiraling (Figure~\ref{fig3}(D-E), see SV 1-4, SV 11-12 and Section D of the Supplementary sheet for more). As the droplet interacts with the translating and rotating vortex ring, the combined action of shear and pressure gradients continuously stretches it both azimuthally and along the direction of propagation. The azimuthally stretched liquid wraps around the core to form a spiral, which is further elongated by the azimuthal pressure gradient \citep{Jha_Govardhan_2015}, producing thin ligaments (see SV11). Once the ligament diameter becomes sufficiently small, capillary instability triggers Rayleigh-Plateau breakup of the ligament into daughter droplets. Simultaneously, the liquid stretched in the translational direction also fragments into fine ligaments and droplets. Many of these daughter droplets remain entrained within the vortex field, where continued local inertial loading can exceed capillary resistance provided by daughter droplets, leading to secondary atomization and further fragmentation (SV 12, 13).

 When the viscosity ratio is much larger than unity, the influence of surface tension becomes less important and break-up rate of droplets decreases, leading to formation of long stable ligaments \citep{Roccon2017,Deike_2023} as in also seen here for olive oil ($(\mu_d/\mu_c)_\mathrm{olive} = 70$). For JetA with $(\mu_d/\mu_c)_\mathrm{jetA} = 1.22$, it is observed that (SV 14 and 15) that interactions with vortex lead to atomization of droplets without long stable threads. The viscosity ratio along with surface tension being very low supports faster break-up with minimal stretching and ligaments unlike seen in the case of olive oil. The fluid properties of spilled live oil during DwH were $(\mu_d/\mu_c)_\mathrm{live oil} < 1$ and $\sigma \approx 37 mN/m$ \citep{Zhao2015} which is expected to show behavior similar to JetA with minimal ligament formation stabilized through surface tension. In another study \citep{Balaji_PRL}, addition of dispersant resulted in order of magnitude reduction in surface tension ($< 1 mN/m$) of the crude oil with viscosity ratio $\sim O(10^1)$ producing fine micron sized ligaments during break-up. The role of low density of JetA is not explored in the present study but it is expected to be disadvantageous during transportation with negligible affect in droplet break-up.

 Interaction of vortex ring with D-II and D-III produces a slightly different phenomenon since the process is mostly dominated by transport rather than atomization like in the case of D-I. The vortex when passes through the droplets creates hollow region which corresponds to the high vorticity zone of the vortex \citep{Avni2022} (See SV 15), also depicted in figure~\ref{fig3}(C)). The passage of the high speed vortex creates a low pressure zone that sucks in the smaller droplets as the vortex exits the cluster. This creates a significant number of droplets to trail the vortex in the wake which are released at a later stage. Hence, the initial interaction only pulls the droplets inside the vortex structure (the region of influence) that settles near the core as time progresses.

 The initial droplet distribution sizes ($d_d$) termed as available droplets ranges from $<0.15 mm$ to $\sim 6mm$  which is relevant to what is predicted for DwH spillage without the usage of dispersant (ranges between 1-10mm) \citep{North_2015,Zhao2015}. However, the ambient fluid properties like temperature along with other rheological properties will also play a crucial role in deciding the nature of the interaction \citep{Zhao2015}. Hence, without accurate experimental setting it is difficult to predict accurately the fate of the droplets. Nevertheless, the process of elongation, roll-up and breakup described above can be extended to real life scenarios and holds significant potential to be able to produce fine dispersions of droplets. 

Post trapping, the droplets are successfully transported for distances $z^*> 50$ especially for higher $Re_\Gamma$ as can be observed from figure~\ref{fig3}(B) and the supplementary videos (refer SV 6, 7, 8 for D-I, SV 9(a-b) for D-II and SV 10(a-b) for D-III). The $\Gamma$ of the vortex ring far downstream ($\sim z^*> 50$) without the droplets become significantly stable (figure~\ref{fig2}(O)). This is hypothesized to be further supported by the presence of droplets of smaller sizes for vortex with higher strength that acts to minimize the azimuthal instabilities and small scale perturbations in the core consequently making the droplet laden vortex stable for long transports. It is observed that a well occupied ring is more stable than an ill-occupied ring. This is sensible because sparse or asymmetric filling of the ring with oil will create unbalanced forces and moments leading to distortion of the ring at later stages. To elucidate the role of the dispersed phase in vortex ring stability, further focused experiments are required, presenting an interesting avenue for future investigation. More results are provided to understand the transport ability of the vortex ring in further sections.

\subsection{Droplet distribution through the journey}\label{subsec2.3}

\begin{figure}
\centering
\includegraphics[width=1\textwidth]{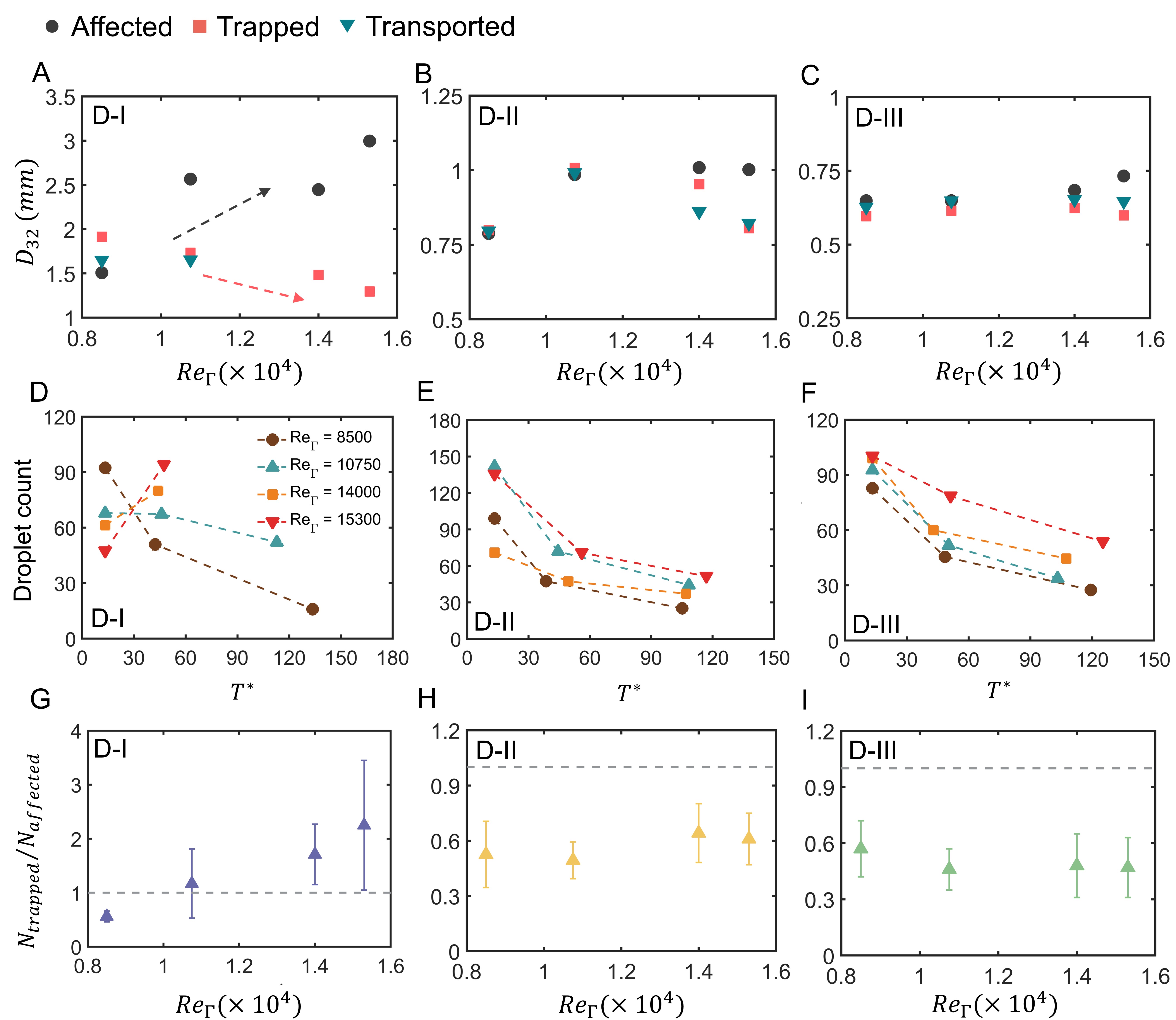}
\caption{(A-C) $D_{32}$ trends of affected, trapped and transported droplets for D-I, D-II and D-III with $Re_\Gamma$. The black dashed arrow in (A) indicates that larger droplets are strongly influenced by stronger vortex. A decrease of $D_{32}$ values marked by red dash imply an increase in atomization with increase in vortex strength (D-F) Average droplet count plotted with normalized time ($T*=tU_\mathrm{conv}/d_\mathrm{pipe}$) for D-I, D-II and D-III (D-F). Trends for ratio of droplet count of trapped to affected droplets for for D-I, D-II and D-III with $Re_\Gamma$. The error bar represents the standard deviation for respective data set.}
\label{fig4}
\end{figure}

We track and estimate the droplet distribution and their respective Sauter mean diameter ($D_{32}$) throughout its journey passing through different stages. Figure~\ref{fig4}(A-C) presents the $D_{32}$  values for affected, trapped and transported droplets for different droplet distributions with $Re_\Gamma$ values (check section E in the supplementary sheet for corresponding probability distribution functions). The $D_{32}$ values plotted here in figure~\ref{fig4}(A-C) is obtained by combining all the runs for each distribution and corresponding $Re_\Gamma$. For D-I distribution where we observe a blend of atomization and transport, a smaller trapped $D_{32}$ would mean more atomization or ejection of larger droplets whereas a larger trapped $D_{32}$ value would imply less atomization and ejection of smaller droplets. Figure~\ref{fig4}(D-F) exhibits the average droplet count as obtained at different stages plotted with $T^*$. An increase in the number of droplets from first to second stage would mean generation of droplets through atomization whereas decrease in the number of droplets from second to third stage would clearly indicate ejection. A decrement from first to second stage includes droplets shed in the wake and the untrapped droplets as well. Since, the interaction phenomenon involves both atomization and transport, it is difficult to exactly bring out the statistics through isolated single indicator. Hence, we plot the $D_{32}$, droplet count and the trapped to affected ratio (in figure~\ref{fig4} (G-I)).

For D-I shown in figure~\ref{fig4}(A), the strongest vortex affects larger droplets the most as reflected by the higher affected $D_{32}$ values for $Re_\Gamma$ = 15300 compared to the lower $D_{32}$ for $Re_\Gamma$ = 8500. The weakest vortex produced in the study corresponding to $Re_\Gamma=3250$ succumbs during the interaction due to high drag and barely affect them (hence, not shown here). It is noteworthy that trapped droplets (captured by camera 2) mostly includes the atomized droplets for $Re_\Gamma\geq 10750$. The $D_{32}$ values corresponding to trapped droplets decreases with increase in the $Re_\Gamma$ value indicating better atomization. This is well captured in figure~\ref{fig4}(D) where for $Re_\Gamma= 10750$ we see similar number of droplets from first to second point in time since atomization does not happen very effectively for this case. Whereas above $Re_\Gamma=10750$, a significant rise in the number of droplets is observed which is a sign of atomization. For $Re_\Gamma=8500$, we see that the vortex is able to trap larger sized droplets better because of lower $St$ but with minimal atomization, a decrease in the $D_{32}$ for transported droplet indicate high ejection in its journey. This is clear from the sharp dip in the droplet count in figure~\ref{fig4}(D) for $Re_\Gamma=8500$.  For $Re_\Gamma=10750$ the transported droplets exhibits similar $D_{32}$ as trapped droplets implying almost no ejection or atomization on its ways till the target which can be corroborated from figure~\ref{fig4}(D). A similar trend is expected for higher $Re_\Gamma$ which could not be measured due to extreme clustering of droplets (refer to SV 8).

For D-II and D-III (figure~\ref{fig4}(B-C)), the gap between the affected and trapped $D_{32}$ values diminishes at lower $Re_\Gamma$ with some atomization and ejection occurring only in cases with $Re_\Gamma\geq 14000$. A significant decrease in the number of droplets is observed for all cases D-II and D-III. This simply means that smaller droplets are not trapped very efficiently. From figure~\ref{fig4}(B-C) also a lower $D_{32}$ for trapped droplets at $Re_\Gamma\geq 14000$ indicates that the vortex misses the droplet. For D-I, the effect of high $St$ number manifests in the form of atomization (will be discussed in section~\ref{subsec2.5}). However, for D-II and specially D-III, a larger $St$ because of very small flow time scale at high $Re_\Gamma$ results in lower trapping. Although counter-intuitive, for better transport, the ring velocity must be optimized to obtain smaller $St$ as much as possible.

For more insights into atomization, we plot the average of the ratio of droplet count of trapped to affected droplets ($N_{trapped}/N_{affected}$) in figure~\ref{fig4}(G-I). Then, a value $>$ 1 must indication generation of droplets and $<$ 1 would mean loss of droplets (i.e. ejection). Higher value of this ratio (above 1) would imply more atomization and lower value would mean more ejection. From figure~\ref{fig4}(G) it is evident that atomization improves remarkably with increase in the vortex strength. For $Re_\Gamma=8500$ the ratio remains much below 1 signifying no atomization. Large error bars indicates the stochastic nature of the interaction process and subsequently, the output of the atomization. Since, breakup of droplets must also depend on their relative orientation with the vortex core which has not been considered, the stochasticity in this data is expected. For D-II, we see the values are slightly higher for higher $Re_\Gamma$ but $<$ 1. Since we are plotting the average values of the run, a value $<1$ only means that in majority of the cases, the vortex is not able to break the droplet or the number of droplets atomized is minimal. However, in the case of D-III, we see lower values of ratio at higher $Re_\Gamma$ because of weaker transport at high $St$.

In this sense, our observations suggest that atomization, somewhat counterintuitively, facilitate transportation of larger droplets as well. As shown in figure~\ref{fig5}(M, N), the engulfed droplets produced through atomization remain within the region of strong vortex influence, close to the vortex core, and are therefore transported efficiently with the ring. In contrast, smaller droplets that do not undergo atomization are initially displaced away from the vortex and subsequently drawn toward the centerline, requiring them to traverse the vortex field before becoming entrained. This pathway appears to be comparatively less favorable for transport than the direct engulfment of atomized droplets.

%A more systematic investigation using a more monodisperse range of initial droplet sizes would help isolate these effects and provide a more rigorous comparison between the two transport pathways.

\subsection{Volumetric Efficiency}\label{subsec2.4}

We define three volumetric efficiencies ($\eta)$: Trapping efficiency ($\eta_\mathrm{trap}$), transport efficiency ($\eta_\mathrm{transport}$) and net efficiency ($\eta_\mathrm{net}$) as follows:
\begin{equation}
\eta_\mathrm{trap} = \frac{\text{Volume of trapped droplets}}{\text{Volume of affected droplets}}
\end{equation}

\begin{equation}
\eta_\mathrm{transport} = \frac{\text{Volume of transported droplets}}{\text{Volume of trapped droplets}}
\end{equation}

\begin{equation}
\eta_\mathrm{net} = \eta_\mathrm{trap}\times\eta_\mathrm{transport}
\end{equation}

\begin{figure}[!h]
\centering
\includegraphics[width=1\textwidth]{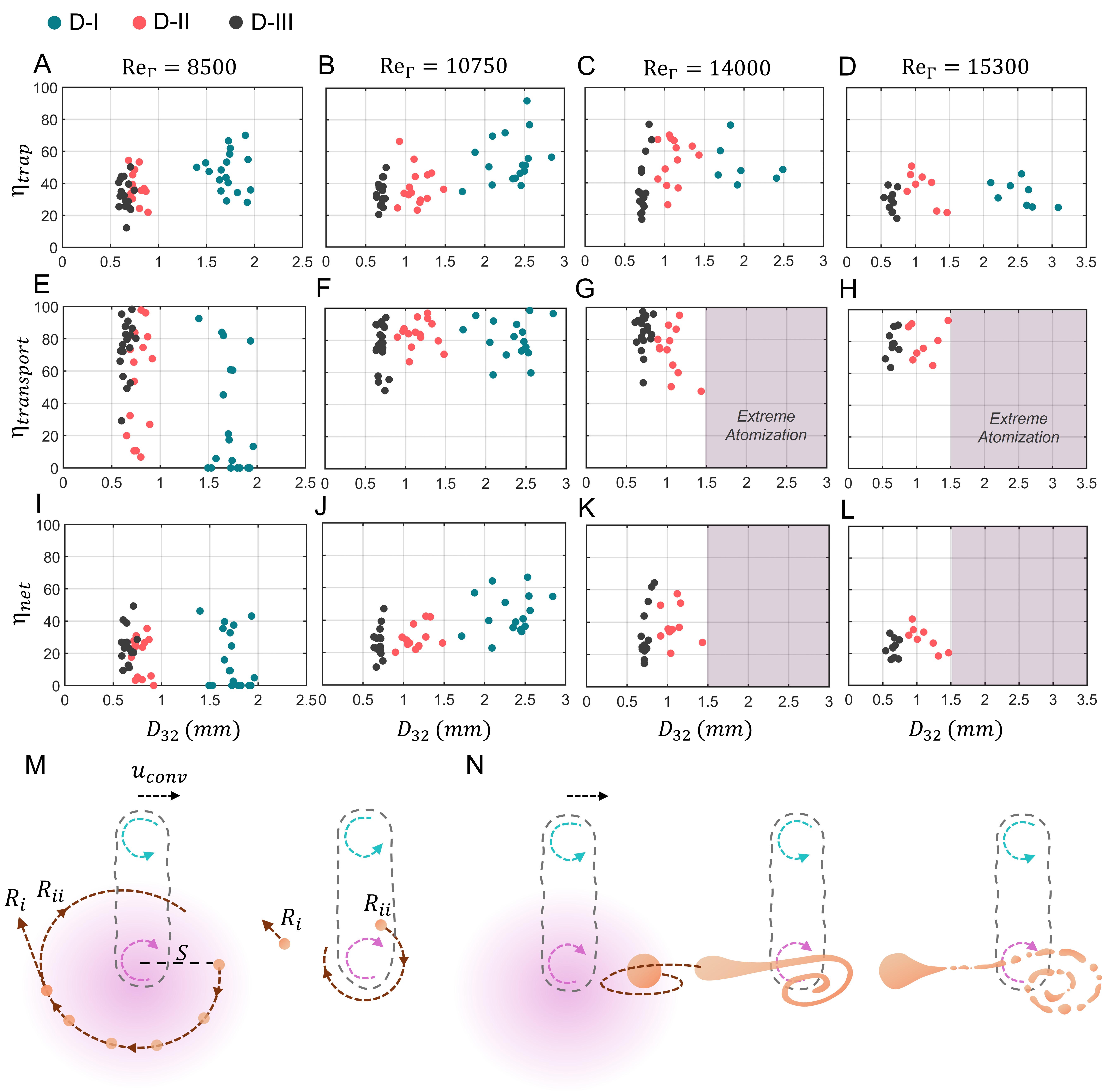}
\caption{The scatter plots representing (A-D) volumetric trapping efficiency ($\eta_{trap}$) (E-H) $\eta_{transport}$ and (I-L) $\eta_{net}$ values with the $D_{32}$ values of the corresponding run (available droplets) for different $Re_\Gamma$. The dotted line represents the slope of all the scatter point. The highlighted zone represents the atomization zone where the data could not be processed. A schematic (not to scale) depicting two scenarios where a (M) small and (N) large droplet is seen to interacting with the lower core of a vortex ring. The pink zone represents the region of influence or the induced field of vortex. $R_i$, $R_{ii}$ represents route $i$ and $ii$ that a droplet can follow. $S$ marks the distance between the core center and the droplet.}
\label{fig5}
\end{figure}

Figure~\ref{fig5} outlines the trends for all the three efficiencies and their averages for different vortex $Re_\Gamma$ sorted as per the $D_{32}$ obtained for each run. For higher $Re_\Gamma\geq14000$ the interaction produces dense clusters of droplets that are challenging to resolve using the present image processing technique (SV 6-8). Following the interaction, the vortex ring often becomes unstable and tilts slightly, providing an oblique three-dimensional view from Camera 2 that enables the trapped droplets to be identified and quantified. As the droplet laden vortex ring continues to propagate, it gradually regains stability, causing the droplets to overlap in the side view and rendering their detection impractical. Therefore, for $Re_\Gamma \geq 14000$, we report only those cases that permit reliable image processing.

The scatter suggest that the $\eta_\mathrm{trap}$ for larger droplets show significant dispersion which is expected since the volumetric efficiency $\propto$ $d_{d}^3$ making it sensitive to larger sizes of droplet (figure~\ref{fig5}(A-D)). Cases with smaller $D_{32}$ exhibit much more coherency in the scatter since they are comprised of smaller droplets. Another reason for higher dispersion of data at high $Re_\Gamma\geq10750$ for larger droplets is atomization. Since for cases with $Re_\Gamma>8500$, the $\eta_\mathrm{trap}$ includes the contribution of both atomization and transport, the effect of atomization being stochastic makes the spread large. Although the plots shown in figure~\ref{fig5}(A-C) suggest higher $\eta_\mathrm{trap}$ for D-I distribution, it is arguable that these value depend strongly on the definition of affected droplets. Since, many larger droplets, by virtue of their inertia, will not satisfy the displacement criteria (equation~\ref{eqn 1}) leading to their exclusion from the sampled population of affected droplets and subsequently contribution to the $\eta_\mathrm{trap}$. A droplet-size dependent definition of effective displacement providing a refined cohort of affected droplets is expected to reduce the $\eta_\mathrm{trap}$ for D-I and maybe D-II. As discussed in section \ref{subsec2.2}, the hollow region created by vortex compels the droplet to orbit around the core region due to its induced field (as depicted in figure~\ref{fig5}(M)). Due to this, most of the droplets then never catch up with the vortex also leaving a trail of smaller droplets in the wake as has also been reported \citep{Avni2022}. This is further dependent on the local $St$ for the small sized droplet. The distance $S$ between the vortex core center and the droplet becomes relevant especially in case of interaction with smaller droplets. A larger $S$ means that the droplet is under weaker influence and may choose $R_i$ route as depicted in figure~\ref{fig5}(M) and vice-versa in case with lower value of $S$. The bifurcation point between the two routes, $R_i$ and $R_{ii}$, can occur anywhere within the rotating flow field, prone to perturbation/turbulence and is arbitrary shown behind the vortex. However, owing to the negative pressure induced by the translating vortex, the droplets predominantly converge toward the vortex from behind, along its wake. Whereas in case of D-I, the larger droplets do not get significantly displaced, as a result, a part of it stretches, leading to atomization (see figure~\ref{fig5}(N)). The newly formed droplets are already in the field of influence of the vortex ring which gets easily convected unlike in earlier case (D-III) where the droplet needs to penetrate inside the vortex field after being pushed. Moreover, the $St$ achieved even for smaller size droplets with slowest moving vortex ring remains near to unity indicating difficulty of droplets to follow the flow.

The $\eta_\mathrm{transport}$ (figure~\ref{fig5}(E-H)) for the $Re_\Gamma=8500$ ranges from $\sim 0 - 95\%$ since the droplets doesn't always get carried till the target leading to $\eta_\mathrm{transport}=0\%$ and subsequently a $\eta_\mathrm{net}=0\%$. For higher $Re_\Gamma$, the majority of the cases show efficiency values $> 50\%$ with the average suggesting vortex ring with $Re_\Gamma = 10750$ and $14000$  to be best performers. The observations for $Re_\Gamma\geq14000$ (SV 8) suggest that vortex rings at higher $Re_\Gamma$ transport a substantially larger quantity of droplets, with the transported population being predominantly composed of smaller droplets. The net efficiency (figure~\ref{fig5}(I-L)) that essentially means the final transported volume of droplets to the initially affected volume shows coherent scatter for $Re_\Gamma = 10750$. For practical purposes, if atomization is the priority, higher $Re_\Gamma$ will give better results due to increased shearing that leads to better atomization (will be discussed in the next section). Whereas for transportation, it is critical to understand the droplet size distribution that needs to be transferred and then decide on the $Re_\Gamma$ value based on assessing the $St$. Furthermore, a carefully curated mix of vortex strengths can also be employed to fine tune the balanced implementation of simultaneous or sequential atomization and transport.

\subsection{Response and Atomization criteria}\label{subsec2.5}

The Stokes number ($St$) in general gives an estimate of how faithfully dispersed phase particle (here droplet) follow the continuous phase fluid (here water). The lower the value of $St$, the faster is a droplet's response to an impulsive flow. For the present problem, we can define $St$ in two ways firstly, using the convection time scale ($d_\mathrm{ring}/u_\mathrm{conv}$)\citep{deAquino2025} and, secondly using the core rotation time ($2\pi r_\mathrm{core}^2/\Gamma$) \citep{Jimenez1996}.The primary requirement for the vortex ring is to first capture the droplets before subsequently transporting or atomizing them. Here, trapping does not imply that the droplets have fully migrated into the vortex core, a process that typically occurs over a longer time scale. Instead, trapping refers to the initial capture of droplets by the vortex flow, that may manifest as a sufficient/significant displacement of droplet from its initial trajectory or an initial orbital motion around the moving core. Accordingly, the following criterion is sufficient to characterize if droplets would respond to the flow:

\begin{equation}
\frac{d_\mathrm{ring}}{u_\mathrm{conv}} \geq \frac{\rho d_d^2}{18 \mu_c}
\label{eqn 4}
\end{equation}

\begin{figure}[H]
\centering
\includegraphics[width=1\textwidth]{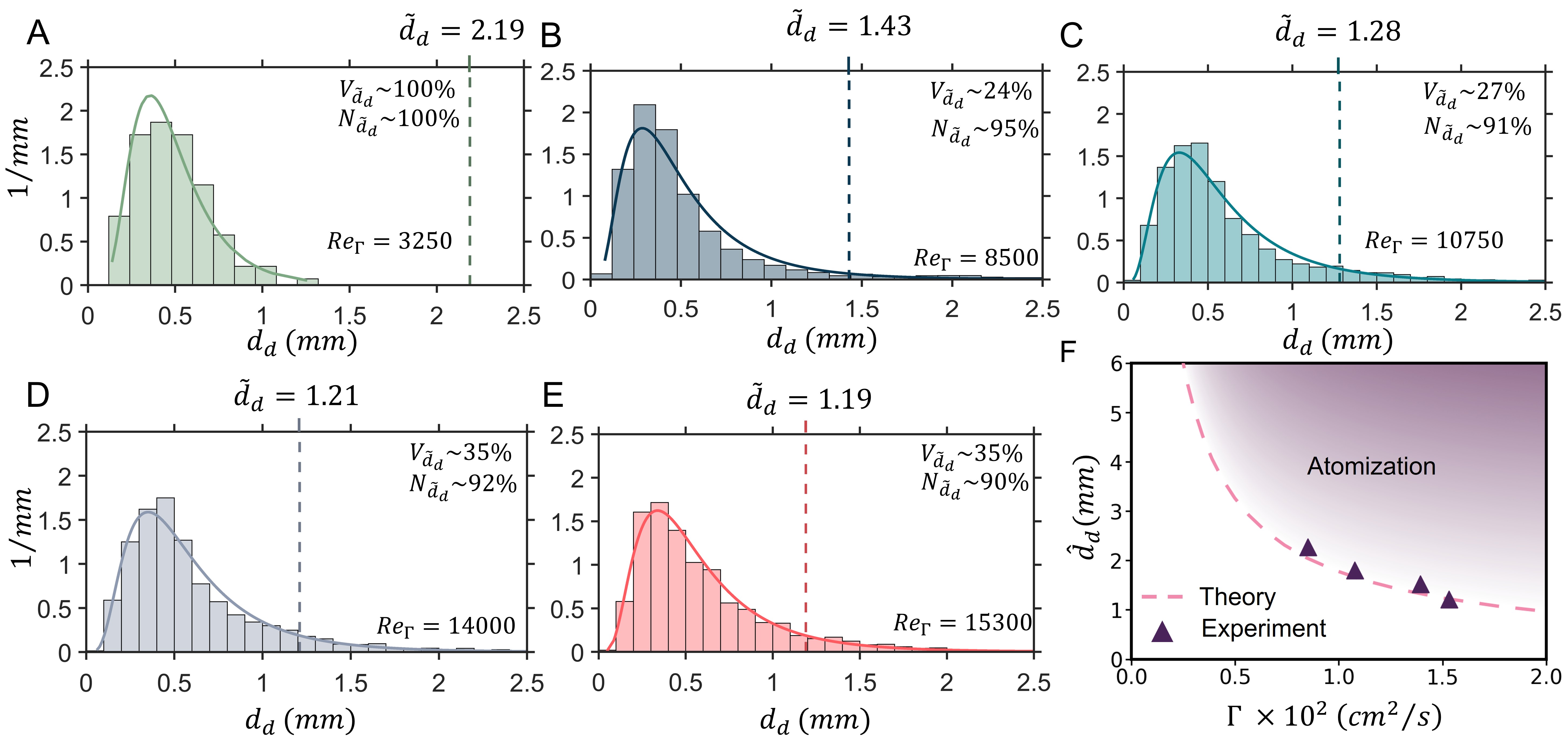}
\caption{(A-E)Probability distribution function of the trapped droplets for each $Re_\Gamma$ fitted with log-normal distribution. The dotted line indicates the threshold diameter for trapping of droplets ($\tilde d_d$) obtained from equation~\ref{eq 5}. $V_{\tilde d_d}$ represents the volume percentage of droplets and $N_{\tilde d_d}$ number percentage of droplets below the obtained threshold that is predicted by equation~\ref{eq 5}. (F) The variation of the threshold diameter for atomization obtained through solving equation~\ref{eq 10} (details in section H supplementary sheet) and experiments. The pink zone is the region where atomization is expected to occur.}
\label{fig6}
\end{figure}

The RHS of the above equation represents the response time of the droplet (see section F in supplementary sheet). This yields the condition to obtain the threshold diameter $\tilde d_d$ for getting displaced:

\begin{equation}
\tilde{d_d} = \left(\frac{18\mu_c d_\mathrm{ring}}{\rho_d u_\mathrm{conv}}\right)^{1/2}
\label{eq 5}
\end{equation}

then for $d_d \leq \tilde d_d$ the droplets should follow the path of the vortex ring. The droplets fulfilling the above criteria may or may not get trapped in the vortex ring. This significantly depends on the relative orientation of the droplet with respect to the vortex center and core. The displaced non-atomized droplets must return to the vortex field after getting initially displaced as shown above in figure~\ref{fig5}(M). Thus, a more rigorous way to achieve droplet trapping criteria would involve solving the Maxey-Riley equation \citep{Maxey_Riley} to get the exact droplet path. However, simultaneous experimental measurements of the fluid flow field with swarm of droplets remains a challenging task. Figure~\ref{fig6}(A-E) depicts the result from the above condition (equation~\ref{eq 5}) overlaid on the probability distribution functions (PDF) for the trapped droplets. It is to be noted that the PDFs for the trapped droplets include the atomized droplets as well which is a consequence of trapping itself. The overall trend suggests that with increasing $Re_\Gamma$ the $\tilde d_d$ decreases which is expected since $u_{conv}$ is a strong function of the vortex $Re_\Gamma$ which appears in inverse relation with the $\tilde d_d$. The criteria predicts the number of displaced/trapped droplets ($N_{\tilde d_d}$) reasonably well even for highest $Re_\Gamma$. To obtain the volume prediction ($V_{\tilde d_d}$) falling under the criteria i.e., affected population (left side), the following formulation is used that even allows to assume a power law form of the tail of the PDF such as $p(d_d)=\alpha d_d^{\ -\beta}$:

\begin{equation}
    V_{\tilde d_d} = \left[1-\frac{\int^{d_{max}}_{\tilde{d_d}} d_d^3 \ p(d_d)dd_d}{\int^{d_{max}}_{0} d_d^3\ p(d_d)dd_d}\right]\times100
    \label{eqn Vdd}
\end{equation}

%\begin{equation}
 %   V_{\tilde d_d} =\left[1- \frac{\alpha}{\forall(4-\beta)}(d_\mathrm{max} \!\!\!\!\!\!\! ^{4-\beta}-\tilde d_d \! ^{4-\beta}) \right] \times 100 
 %   \label{eqn power}
%\end{equation}

We also present the same analysis (as in equation~\ref{eqn 4}) with the core rotation time (refer to Section F in supplementary sheet) and observe that the balance using the convection time scale performs significantly better than the core rotation time. This is because realistically, the zone of influence for the droplets due to a traveling vortex ring is much larger compared to its core size and the droplets doesn't penetrate the core instantly. Instead, most of the time they convect with the vortex (including in its wake which can be observed from SV 5) and gradually occupy the core. This happens based on the dominance between the  centrifugal forces and the suction created in the core. A larger centrifugal force would tend to eject the droplet away whereas the the pressure gradient across the vortex core would try to pull the droplet in. All this makes sense only when the droplet response time is of the order of flow time scale. It is expected that the particle spiral in towards the vortex core which occurs over larger time. The convective time scale being larger than the core rotation time (see section G in supplementary sheet) therefore, gives a better criteria. 

Furthermore, similar behavior of dispersed phase is expected only when the value of $St$ number is $\ll1$. A larger $St \geq 1$ results in a lag in the particle's velocity compared to the continuous flow. A vortex ring always induces a velocity field around it with a wake which is also capable of stretching and atomizing larger droplets that exhibit larger $St$. Under this condition, a larger droplet is subjected to larger lag and hence velocity difference with respect to the continuous phase resulting in significant shearing from the external vortical field and stretching, leading to atomization. Following this, it can be understood that for atomization, the shear forcing on the droplet surface must be higher than the resistive surface tension force from the droplet. To formulate this, we assume that the relevant velocity scale of the vortex to be $U_{ref}={\Gamma}/{2\pi r}$ where $r$ is the radial coordinates of the vortex core. Then, the local shear rate across the droplet can be expressed as $\dot\nu=\partial U_{ref}/\partial r$. The elongation of the droplets from one end (as observed in figure~\ref{fig3} and section D of supplementary sheet) arises from the velocity gradient across the droplet, which generates differential advection and causes it to stretch axially along the vortex. This velocity difference across a droplet ($\Delta u$) can be approximated to be $\dot{\nu} d_d$. The shear force on a droplet inside a vortical field can then be evaluated as $F_\mathrm{shear}={1}/{2} C_d\rho_c (\Delta u)^2A_p$. Here, the drag coefficient ($C_d$) is estimated through the Schiller–Naumann drag coefficient model \citep{schiller1933} and $A_p$ refers to the characteristic area of the droplet. For atomization to occur, the shear force term must exceed the resistive interfacial tension force ($F_\sigma$) of the droplet given as $\sigma (\pi d_d$). Then, the condition $F_\mathrm{shear}\geq C_k \sigma(\pi d_d)$ should hold where $C_k$ is an experimental constant. Solving this at $r = r_{core}$ we get (details in section H of supplementary sheet) threshold diameter ($\hat{d_d}$) as: 

\begin{equation}
\hat{d_d} = C_k\left(\frac{8 \pi^2 \sigma r_\mathrm{core}^4}{C_d \rho_c \Gamma^2}\right)^{1/3}
\label{eq 10}
\end{equation}

\begin{figure}[H]
\centering
\includegraphics[width=1\textwidth]{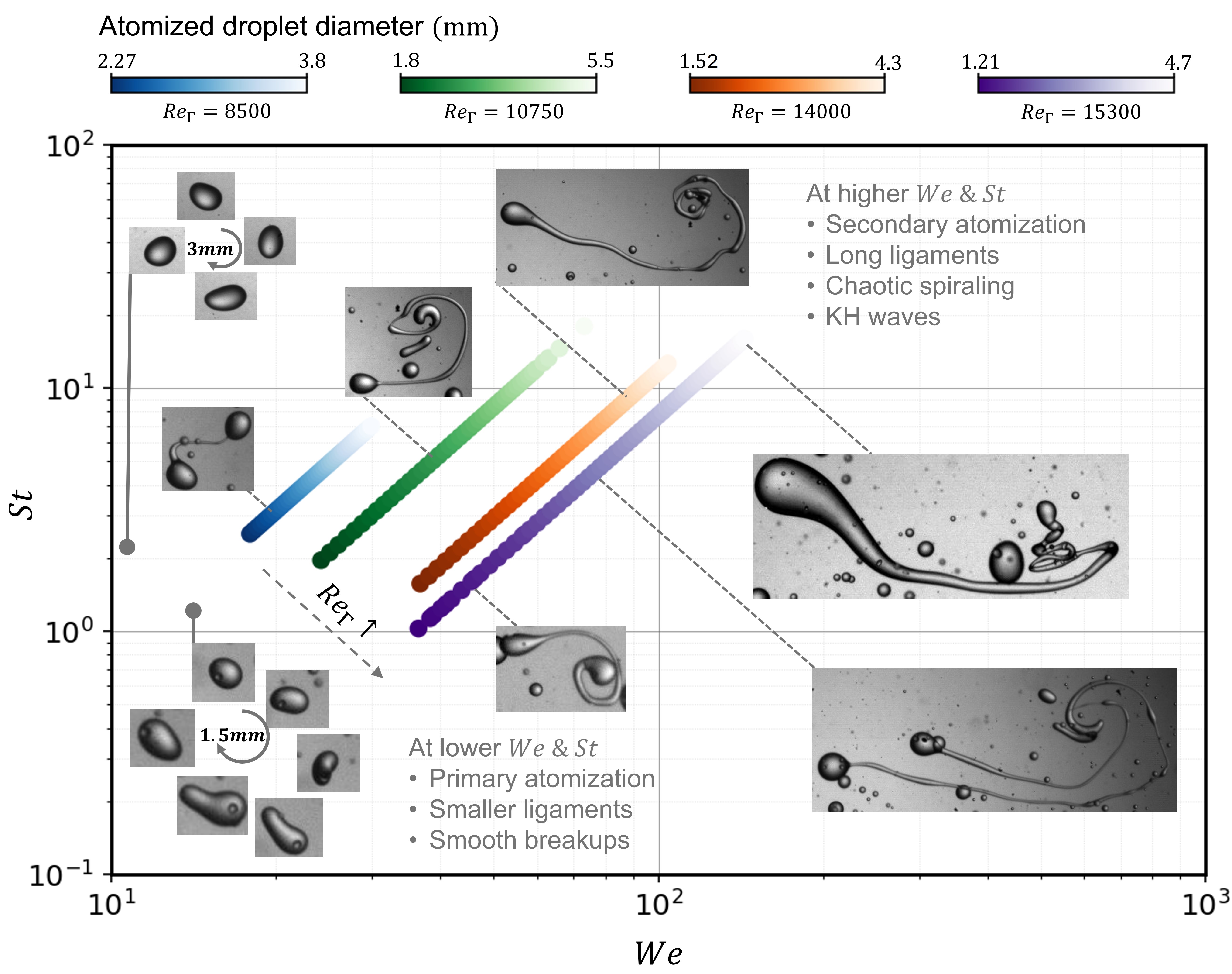}
\caption{A regime map illustrating Stokes number vs Weber number for the range of atomized droplets seen in the present work at different $Re_\Gamma$. Representative images of deformation and break-up are added as insets.}
\label{fig7}
\end{figure}

For $d_d \geq \hat{d_d}$ the droplet will atomize. The results obtained for the $\hat{d_d}$ for atomization through solving the above inequality are plotted in figure~\ref{fig6}(F). Here, we use the area of the full sphere i.e., $\pi d_d^2$ for which the experimental constant ($C_k$) comes out to be unity. However, when we use a projected area of a sphere i.e. $\pi d_d^2/4$ the value of $C_k \sim 0.8$. Since, in our case the whole droplet is submerged within the fluid, the surface area of the sphere seems to be more relevant quantity. The values obtained from equation~\ref{eq 10} shows excellent agreement with the experimentally obtained threshold for atomization. We plot $\hat{d}_d$ as a function of ring circulation in figure~\ref{fig6}(F), aligning the analysis with the present focus of the work, as most studies report the droplet diameter generated during oil spillage rather than the Weber number governing its breakup. However, the formulation can be modified in terms of critical Weber number:

\begin{equation}
We_c = C_i\left(\frac{ \rho_c^2 \Gamma^4}{8C_d \pi^4\sigma^2r_\mathrm{core}^2 }\right)^{1/3}
\label{eq 11}
\end{equation}

where $C_i$ is a constant of $O(1)$. Then, for $We \geq We_c$ we observe atomization. Finally, figure~\ref{fig7} presents the droplet distribution of the atomized droplets as a function of their respective $We$ and $St$ for different values of $Re_\Gamma$. The experimental data plotted in figure~\ref{fig7} can be retrieved from figure~\ref{fig6}(F) where the initial values for each $Re_\Gamma$ indicate the threshold for atomization at respective $Re_\Gamma$. Representative images for droplet deformation and atomization are placed within the plot corresponding to the droplet diameter for a qualitative understanding of the phenomenon. It is evident that the maximum droplet size that can be atomized is considerably smaller at lower $Re_\Gamma = 8500$ than at higher $Re_\Gamma$. This is because larger droplets significantly deform the weaker vortex rings, leading to a loss of ring coherence and a reduction in the energy available for atomization.
Interestingly, the atomization primarily begins when $St\sim1$. As discussed above, a larger $St$ number indicates larger relative drag between the two phases. Given a sharp velocity gradient of the vortex field, this results in shearing and atomization of droplets. The map portrays $St$ number as an indicator for atomization along with $We$ which is traditionally used in the literature. Hence, for soft deforming dispersed phase in an unsteady flow field, $St$ can be a complementary indicator for atomization.

\section{Discussion}\label{sec3}

The present study demonstrates the feasibility of using coherent vortex rings as a chemical free strategy for underwater oil spill mitigation. Unlike chemical dispersants, which fragment droplets by lowering the oil-water interfacial tension, vortex rings promote breakup by increasing the local Weber number through enhanced inertial forcing. This way the the properties of the oil is also preserved which is not the case with dispersants. The resulting smaller droplets provide several potential environmental benefits, including slower ascent, enhanced biodegradation and bioavailability, faster dissolution, and the formation of thinner surface slicks. Using an underwater oil jet interacting with vortex rings spanning five  $Re_\Gamma$, we show that vortex rings can atomize droplets of relevant size range while transporting them over substantial distances ($z^* \geq 50$). We further establish physically motivated criteria for droplet response and atomization based on characteristic time scales and force balances, and demonstrate through single droplet experiments the role of fluid properties on the breakup dynamics.

\begin{figure}[H]
\centering
\includegraphics[width=1\textwidth]{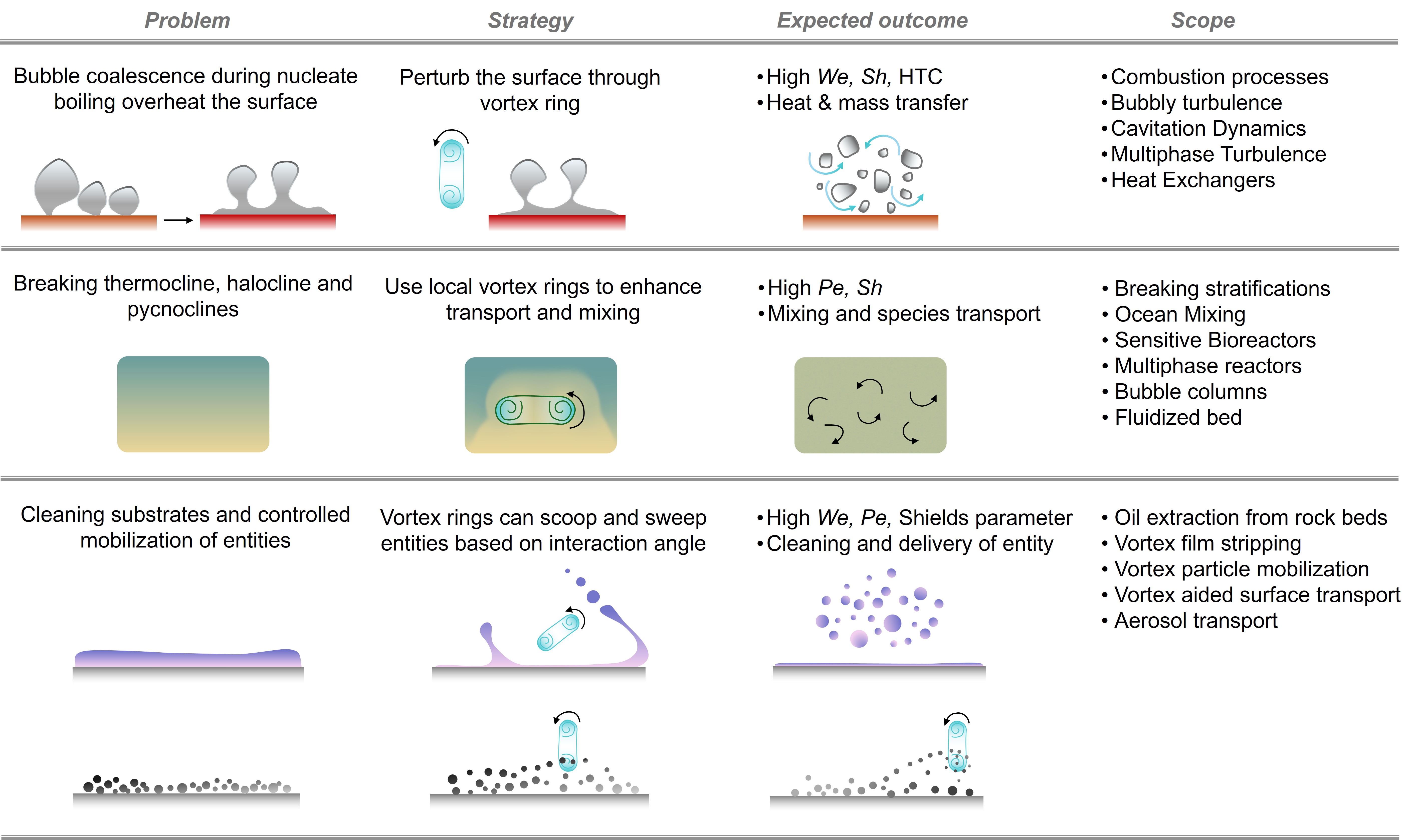}
\caption{Application of vortex ring can be expanded to several aspects of fluid dynamics, transport processes and heat transfer. Here, we show through schematic a few examples of where vortex ring can find applicability in improving efficiency of various processes. In the figure, $Sh$ represents Sherwood number indicating enhancement of mass transfer due to convection relative to molecular diffusion. HTC refers to heat transfer coefficient and a higher HTC means better convective heat transfer. $Pe$ represents the Peclet number that is the ratio of advective transport to diffusive transport. The Shields parameter typically indicates the ratio of fluid forces to the weight of the particle.}
\label{fig8}
\end{figure}

Our results suggest that coherent vortex rings could address two key challenges in underwater oil spill response: reducing the size of buoyant oil droplets without chemical additives and actively redistributing dispersed oil away from ecologically sensitive regions. Although the present work is limited to laboratory scale experiments, the underlying physical mechanisms are expected to remain valid across larger scales because the relevant droplet properties, including size, terminal rise velocity, density difference, viscosity, and interfacial tension, are preserved in real spills. A natural question arising from the present study is whether vortex rings can be scaled to field relevant dimensions, or indeed whether such scaling is necessary for effective oil atomization? Nature itself provides compelling evidence that large, coherent, and remarkably stable vortex structures can exist over large scales \citep{Silver_2006, taddeucci2021volcanic}, demonstrating that vortex rings are inherently scalable. Nevertheless, practical deployment in the ocean would require them to possess sufficient circulation and coherence to withstand ambient turbulence and viscous diffusion. An alternative strategy is to employ an array of synchronized vortex ring generators that produce interacting rings from multiple directions. This approach could establish localized regions of intense turbulence and mixing, thereby enhancing droplet breakup similar to what has been proposed recently \citep{Matsuzawa2023}. Future work should therefore focus on validation and development of strategies for employing the presented idea. 

The transport and atomization capabilities of vortex ring could be leveraged across a much wider range of applications, yet they remain surprisingly underexplored. Figure~\ref{fig8} schematically illustrates several broader possibilities, highlighting the potential of vortex rings as a versatile tool for diverse problems in fluid dynamics and transport processes.

 \section{Methods}\label{sec4}

 \subsection{Generation of vortex ring and oil jet}\label{subsec4.1}
All the experiments were conducted in an acrylic tank of size 30cm x 30cm x 90cm with wall thickness of 8mm. Vortex rings of different strengths are generated through an acrylic pipe of inner diameter ($d_\mathrm{pipe}$) 10mm by an arduino controlled solenoid valve connected to a pressurized water chamber as has been reported in our previous studies \citep{Sharma_Singh_Basu_2021, Jain2023, jain2025Vwall, jain2026EXIF}. Through a separate solenoid valve and pressure regulation unit, a nozzle with orifice diameter of 1.5mm is attached to an acrylic tube generating the olive oil jet at $\sim 13^{\circ}$ from vertical axis (as depicted in figure~\ref{fig2}(A)). The generated droplet size distribution measured form side view (camera 1) has been shown in figure ~\ref{fig2}(C-E). For reference, the front of view of the cluster formed by the oil jet has been presented in section I of supplementary sheet.

\subsection{Shadowgraphy setup}\label{subsec4.2}

Three high speed cameras namely Camera 1 - Photron Mini UX100 Camera 2 \& 3 - Photron Fastcam SA5 were used to capture the overall phenomenon in three parts as shown in figure~\ref{fig2}(A). The recording rates were kept constant at 1000Hz for all the cameras. Image acquisition  for camera 1 using Photron Mini UX 100 high speed was carried out at a spatial resolution of 1280 × 1024 pixels, corresponding to a field of view of approximately 67mm × 52mm. For camera 2 and 3 (Photron Fastcam SA5), the spatial resolution obtained was 1024 x 1024 pixels  with a field of view of $\sim$ 55mm x 55mm. Opposite to each of the cameras, a light source along with diffuser plates were aligned to obtain the shadow of the event. The distance between the cameras were kept such that the different events as segregated (figure ~\ref{fig3}) could be captured. The final target distance to reach by the vortex for transport of droplets was kept at $z^*\sim50$ from the vortex ejection point. However, it was seen that at higher strengths, the vortex was able to transport oil droplets to even larger distances.  For single droplet experiments, two cameras were utilized. One from the side view (Camera 1) and other capturing from front end towards which the vortex traveled.

\subsection{Particle Image Velocimetry}\label{subsec4.2}

Particle Image Velocimetry (PIV) measurements were performed using a dual-pulse high-speed Nd laser (Photonics Industries, 527 nm wavelength, 30 mJ pulse energy). The images were acquired at a recording rate of 1000 Hz. The laser beam was transformed into a planar light sheet of approximately 1 mm thickness using cylindrical optics to ensure uniform illumination of the measurement plane. The laser was set on top of the free surface allowing spreading of the laser into the acrylic tank illuminating the region of interest. The flow was seeded with neutrally buoyant borosilicate glass particles (Sigma-Aldrich) having diameters of 9–13 µm and a density of 1100 $kg/m^3$ as has been done in our previous studies \citep{Sharma_Singh_Basu_2021,Jain2023}. Velocity vectors were computed using a multi-pass cross-correlation algorithm with interrogation window sizes successively reduced from 64 × 64 to 32 × 32 pixels over three passes, while maintaining a 50\% overlap. This resulted in a vector spacing of 0.84 mm in both the streamwise and transverse directions. A B-spline bi-cubic interpolation scheme was employed during the final pass, followed by median and denoising filters to suppress spurious vectors. The processed data was used to obtain the vector field and subsequently identify the vortex core, convection velocity and circulation of the vortex rings using the following relation:

\begin{equation}
 \Gamma = \int_C\omega dA
\end{equation}

where  \textit{C} is the region of interest. For all the cases, a threshold of 10\% of the maximum vorticity was considered to calculate the circulation that sufficiently  eliminated the background noise \citep{jain2025Vwall}.\\
\\

%\subsection{Statistical analysis}\label{subsec3.2}
% Additional details can be found in Methods

\textbf{Acknowledgements}: S.J and S.J.R would like to thank the Prime Minister Research Fellowship (PMRF) for the financial support. S.B. would like to acknowledge the support from the Indian National Academy of Engineering (INAE) Chair professorship.

\textbf{Declaration of interests}: The authors report no conflict of interest.

\bibliography{sn-bibliography}% common bib file

\newpage

%% ==================================================================

\textbf{\large{Supplementary Sheet}
}
\section{A: Rising terminal velocity (\texorpdfstring{$v_d$}{vd}) derivation}\label{sec:terminal}

The terminal velocity is the velocity attained by the rising droplet once the buoyancy force completely balances the drag force on the droplet. Hence, to find the terminal velocity, we equate the buoyancy force with the drag force:
\begin{align}
(\Delta\rho)\,\forall_d\, g &= C_d\,\frac{1}{2}\,\rho_c\, v_d^{2} A, \\
(\rho_c-\rho_d)\,\frac{\pi}{6}\,d_d^{3}\, g &= C_d\,\frac{1}{2}\,\rho_c\, v_d^{2}\,\frac{\pi}{4}\,d_d^{2}, \\
v_d &= \left(\frac{4}{3}\,\frac{(\rho_c-\rho_d)\,d_d\, g}{\rho_c\, C_d}\right)^{1/2},
\end{align}
where $v_d$ is the rising velocity of the droplets, $\forall_d$ and $A$ are the volume and projected area of the droplet, $\rho_c$ and $\rho_d$ are the densities of the continuous and dispersed media, $g$ is the gravitational acceleration, and $C_d$ is the drag coefficient calculated using the following implicit formulation~\cite{Friedman2002}:
\begin{equation}
C_d \approx 0.4 + \frac{24}{Re_d} + \frac{6}{1+\sqrt{Re_d}}, \qquad Re_d = \frac{v_d\, d_d}{\nu}.
\end{equation}

%% ==================================================================
\section{B: \texorpdfstring{$\Gamma_2$}{Gamma2} method for vortex core detection}\label{sec:gamma2}

The use of vorticity fields to detect vortices can lead to erroneous interpretations \cite{Hussain1995}. Since vorticity is computed from velocity gradients, vorticity fields cannot reliably distinguish between regions of pure rotation and regions dominated by shear \cite{Jain2025}. In the present work, we use vorticity contours to calculate the circulation and the $\Gamma_2$ method to deduce the centres of rotating vortices. The $\Gamma_2$ method, as proposed by \cite{Graftieaux2001}, is defined as

\begin{equation}
\Gamma_2(P) = \frac{1}{N}\sum_{S}
\frac{\left(\overrightarrow{PQ}\times\left(\vec{U}_Q-\vec{U}_m\right)\right)\cdot\vec{z}}
{\left\lVert\overrightarrow{PQ}\right\rVert\cdot\left\lVert\vec{U}_Q-\vec{U}_m\right\rVert},
\end{equation}

where the formulation is applied at each point $P$ in the vector field obtained from the PIV data to find the normalized scalar $\Gamma_2$ values. $\overrightarrow{PQ}$ denotes the displacement vector from point $P$ to $Q$ inside area $S$, $\vec{U}_m$ refers to the mean velocity of the area $S$, $\vec{U}_Q$ refers to the velocity vector at point $Q$, and $\vec{z}$ is the unit vector perpendicular to the plane. The threshold for the vortex centre is taken to be 0.75 (i.e., $>2/\pi$) since the structures formed in the downstream region after interaction are chaotic. The consideration of the local velocity vector $\vec{U}_m$ makes the method Galilean invariant, unlike the $\Gamma_1$ method \cite{Graftieaux2001}.

%% ==================================================================
\section{C: Thresholding to identify the affected droplets}\label{sec:threshold}

\begin{figure}[!h]
\centering
\includegraphics[width=\textwidth]{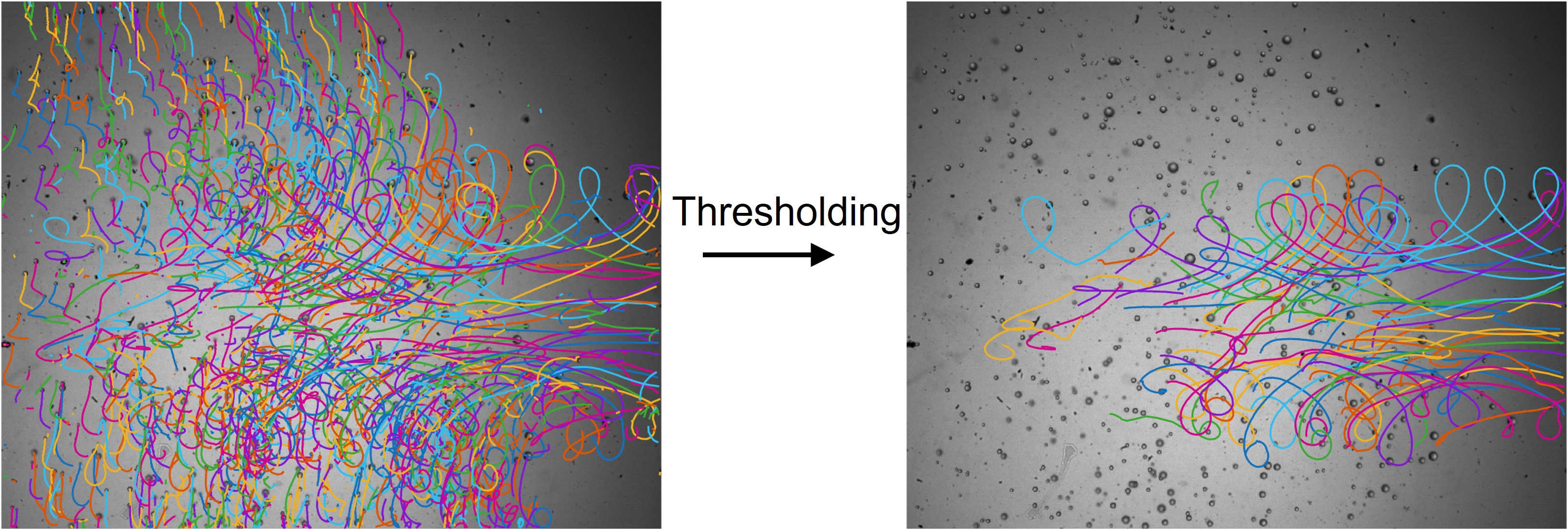}
\caption{Droplet trajectories before (left) and after (right) thresholding to identify the affected droplets.}\label{fig:threshold}
\end{figure}

When a vortex migrates in a fluid, it creates a large zone of influence. Essentially, the magnitude of the velocity (its different components) decreases as we move away from the centre of the vortex in the azimuthal directions. In our case, the interaction or passage of the vortex through the swarm of droplets influences the whole field of view, as can be seen in Fig.~\ref{fig:threshold}. However, it is not practical to consider the whole field as `affected droplets' since this is measurement dependent. Hence, we call a droplet affected only if it has moved a distance equivalent to the core diameter in the direction of the vortex migration. Once this criterion is applied, the code gives the trajectories of only the affected droplets, which may or may not be trapped but are at least displaced.

%% ==================================================================
\section{D: A sequence showing atomization of a droplet}\label{sec:atomization}

\begin{figure}[!h]
\centering
\includegraphics[width=\textwidth]{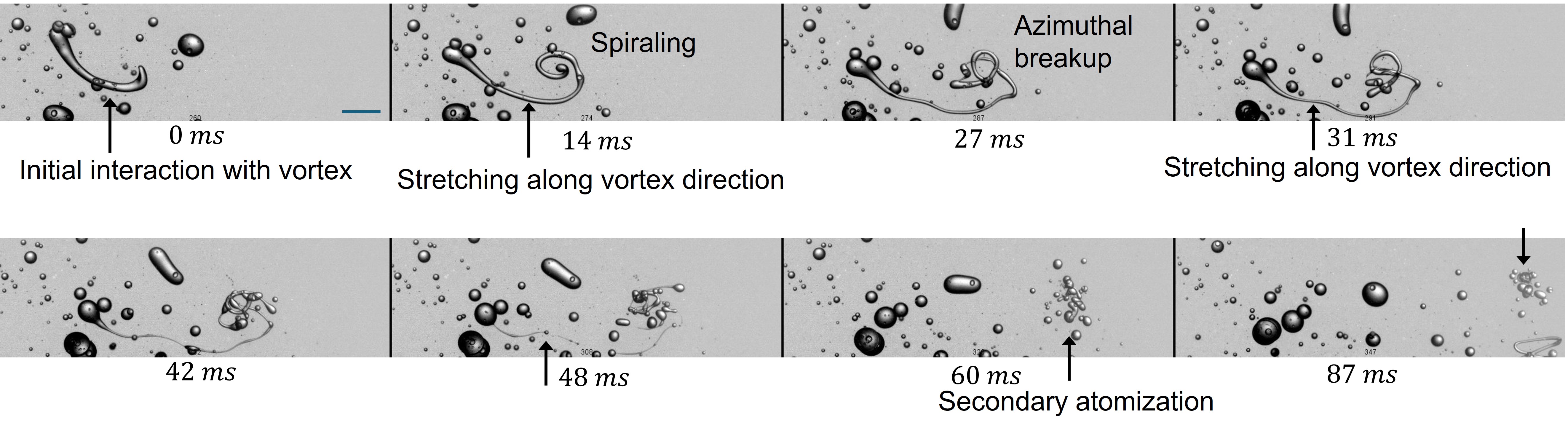}
\caption{Time sequence of a single droplet atomizing under the influence of a vortex ring. The scale bars represent 5\,mm.}\label{fig:atomization}
\end{figure}

Figure~\ref{fig:atomization} depicts time-series snapshots of a droplet being stretched due to the shearing caused in the vortical environment. Stretching in the vortex and azimuthal directions is shown, with subsequent breakup through a Rayleigh--Plateau type instability.

%% ==================================================================
\section{E: Probability distribution functions (PDF)}\label{sec:pdf}

\begin{figure}[!h]
\centering
\includegraphics[width=\textwidth]{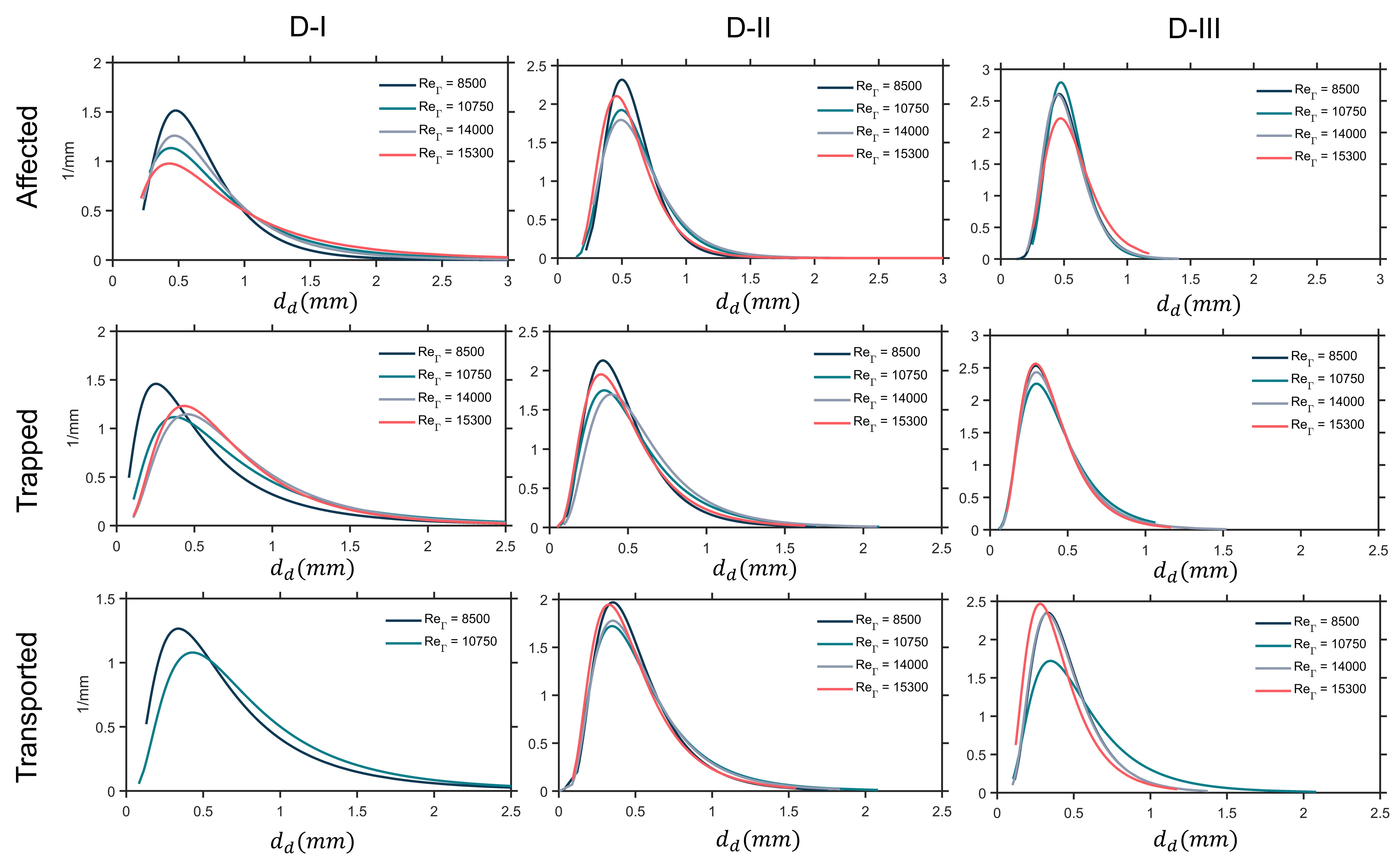}
\caption{PDFs of affected, trapped and transported droplets for the D-I, D-II and D-III initial droplet distributions.}\label{fig:pdf}
\end{figure}

Figure~\ref{fig:pdf} shows the PDFs of affected, trapped and transported droplets for D-I, D-II and D-III at different vortex strengths. A log-normal fit is used, which precisely fits the data set.

%% ==================================================================
\section{F: Response time of droplet calculation and response criteria using core rotation time}\label{sec:response}

Following the arguments put forward in the manuscript, if we consider the flow time scale to be the core turnover time, then for trapping of oil droplets this must be greater than the droplet response time itself. The response time of a droplet can be obtained using Newton's second law for the droplet as
\begin{equation}
m\,\frac{du_p}{dt} = F_{drag},
\end{equation}
where $m$ is the mass of the droplet and $u_p$ represents the droplet velocity. Using Stokes drag, we get
\begin{equation}
m\,\frac{du_p}{dt} = 3\pi\mu\, d_p\,(u_f-u_p).
\end{equation}
Solving the above differential equation, we obtain
\begin{equation}
\ln(u_f-u_p) = -\frac{18\mu}{\rho_p d_p^{2}}\,t + C.
\end{equation}
Using the initial condition $u_p=0$ at $t=0$, we get the final form as
\begin{equation}
\frac{u_p}{u_f} = 1 - \exp\!\left(-\frac{18\mu}{\rho_p d_p^{2}}\,t\right).
\end{equation}
The term $\rho_p d_p^{2}/18\mu$ gives the response time of the droplet. Using the above formulation, we write
\begin{align}
\frac{2\pi r_c^{2}}{\Gamma} &\geq \frac{\rho_d\, d_d^{2}}{18\mu_c}, \\
\tilde{d}_d &\leq \left(\frac{36\pi r_c^{2}\mu_c}{\rho_d\,\Gamma}\right)^{1/2}.
\end{align}

\begin{figure}[!h]
\centering
\includegraphics[width=\textwidth]{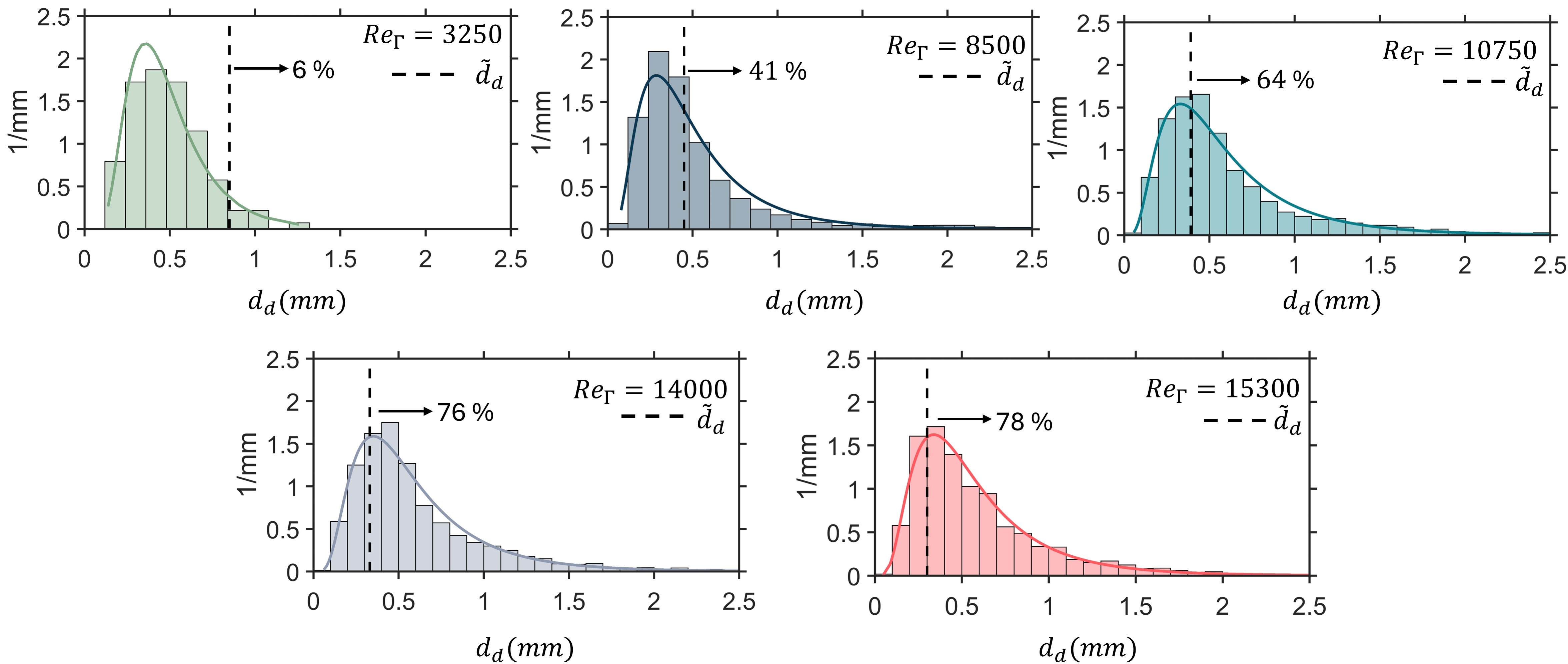}
\caption{Droplet size PDFs at $Re_\Gamma = 3250$, 8500, 10750, 14000 and 15300. The dashed line marks the threshold diameter $\tilde{d}_d$ predicted by the core-turnover response criterion; the percentage indicates the fraction of droplets lying to the right of the line.}\label{fig:response}
\end{figure}

In Fig.~\ref{fig:response}, the region to the right of the dotted line shows the part of the PDF (number of droplets) that is not predicted by the model. The volume predictions are not shown; by observation, these would be very high on the right side of the dotted line. Evidently, the balance between the vortex convection velocity scale and the droplet response time yields better results, as shown in Fig.~6(A--E) of the main manuscript.

%% ==================================================================
\section{G: Comparison between the two natural time scales}\label{sec:timescales}

\begin{figure}[!h]
\centering
\includegraphics[width=0.65\textwidth]{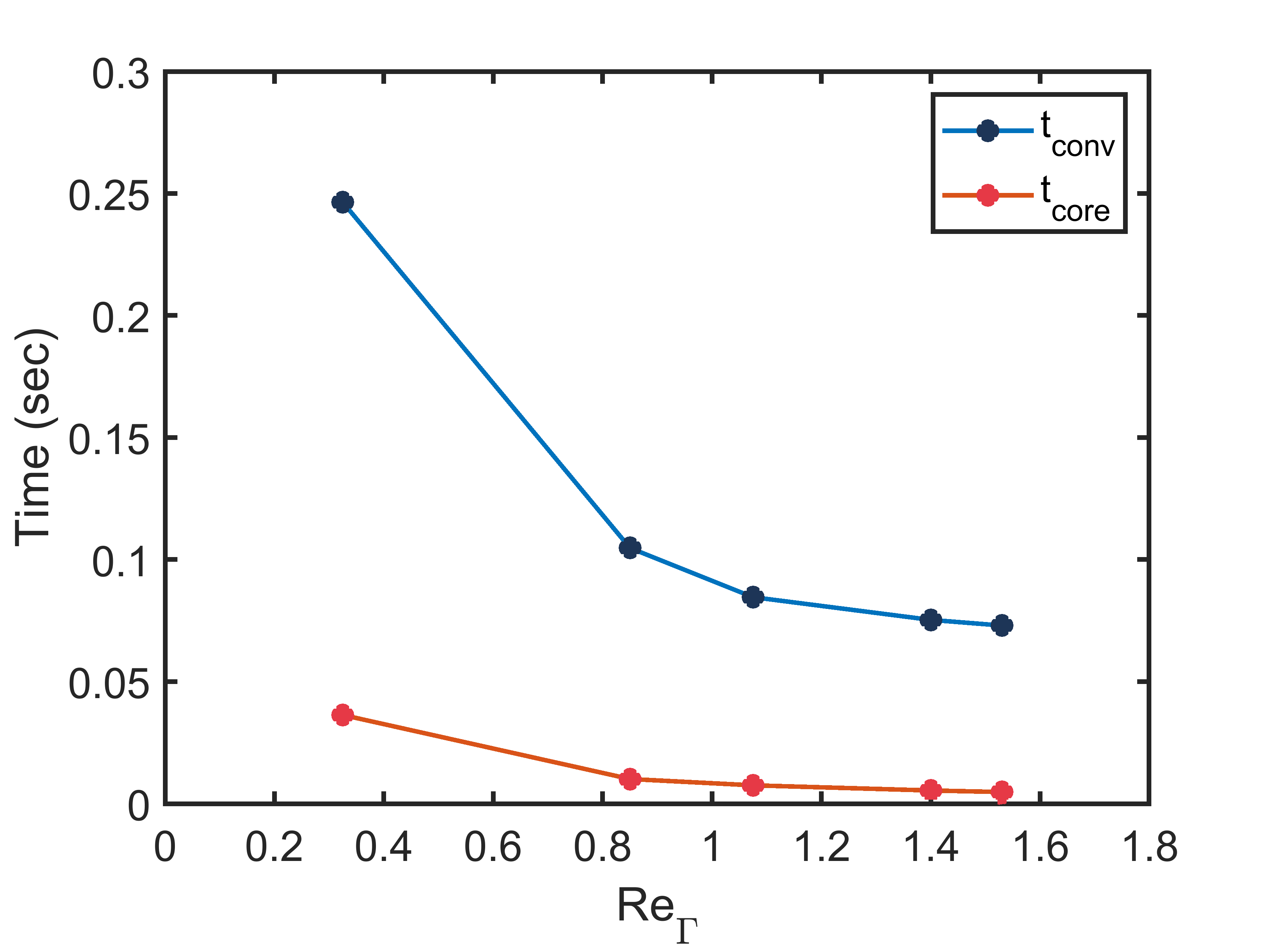}
\caption{Convective time scale $t_{conv}=d_{ring}/u_{conv}$ and core turnover time $t_{core}=2\pi r_{core}^{2}/\Gamma$ as functions of $Re_\Gamma$.}\label{fig:timescales}
\end{figure}

A comparison between the two flow time scales, the convective time scale $d_{ring}/u_{conv}$ (blue) and the core turnover time $2\pi r_{core}^{2}/\Gamma$ (red), is depicted in Fig.~\ref{fig:timescales}.

As can be seen, the core turnover time is much smaller than the convective time scale of the vortex ring. Hence, the trapping criterion performs better with the convective time scale. Using this scale allows sufficient time for the droplets to react and get trapped in the vortex field.

%% ==================================================================
\section{H: Detailed derivation of threshold diameter for atomization}\label{sec:atomthreshold}

We assume the azimuthal velocity of the vortex as follows:
\begin{equation}\label{eq:Utheta}
U_\theta = \frac{\Gamma}{2\pi r_{core}}.
\end{equation}
Then, using Eq.~\eqref{eq:Utheta}, the local shear rate across a droplet of diameter $d_d$ can be written as
\begin{equation}\label{eq:shearrate}
\dot{\gamma} = \left.\frac{\partial U_\theta}{\partial r}\right|_{r_c} = -\frac{\Gamma}{2\pi r_c^{2}}.
\end{equation}
Using Eq.~\eqref{eq:shearrate}, we obtain the velocity difference across the droplet as
\begin{equation}\label{eq:deltau}
\Delta u = \dot{\gamma}\, d_d = -\frac{\Gamma d_d}{2\pi r_c^{2}}.
\end{equation}
Now, the shear force can be expressed as
\begin{equation}\label{eq:Fshear}
F_{shear} = \frac{1}{2}\, C_d\, \rho_c\, (\Delta u)^{2}\,\left(\pi d_d^{2}\right).
\end{equation}
For atomization to occur, the above shear force must overcome the resistive surface tension force of the interface. This can be expressed as
\begin{equation}\label{eq:balance}
F_{shear} \geq F_\sigma.
\end{equation}
Using Eq.~\eqref{eq:deltau}, we get
\begin{equation}\label{eq:balance2}
\frac{C_d\,\rho_c\,\Gamma^{2} d_d^{4}}{8\pi r_c^{4}} \geq \sigma\,(\pi d_d),
\end{equation}
\begin{equation}\label{eq:dhat}
\hat{d}_d \geq \left[\frac{8\pi^{2}\sigma\, r_c^{4}}{C_d\,\rho_c\,\Gamma^{2}}\right]^{1/3}.
\end{equation}
The value of the drag coefficient is determined as follows \cite{schiller1933}. First, we define a relative Reynolds number,
\begin{equation}
Re_r = \frac{\rho_c\,\lvert u_f-u_d\rvert\, d_d}{\mu_c}.
\end{equation}
Then,
\begin{equation}
C_d =
\begin{cases}
0.44, & Re_r > 1000,\\[4pt]
\dfrac{24\left(1+0.15\,Re_r^{0.687}\right)}{Re_r}, & Re_r \leq 1000.
\end{cases}
\end{equation}
This yields a wide range of $C_d$ depending on the droplet sizes. Since a single representative value of $C_d$ is needed, we take $C_d = 0.44$, which corresponds to the mean value of the overall droplet distribution obtained in the study, i.e., 0.59\,mm.

%% ==================================================================
\section{I: Front view of oil jet}\label{sec:jet}

\begin{figure}[!h]
\centering
\includegraphics[width=\textwidth]{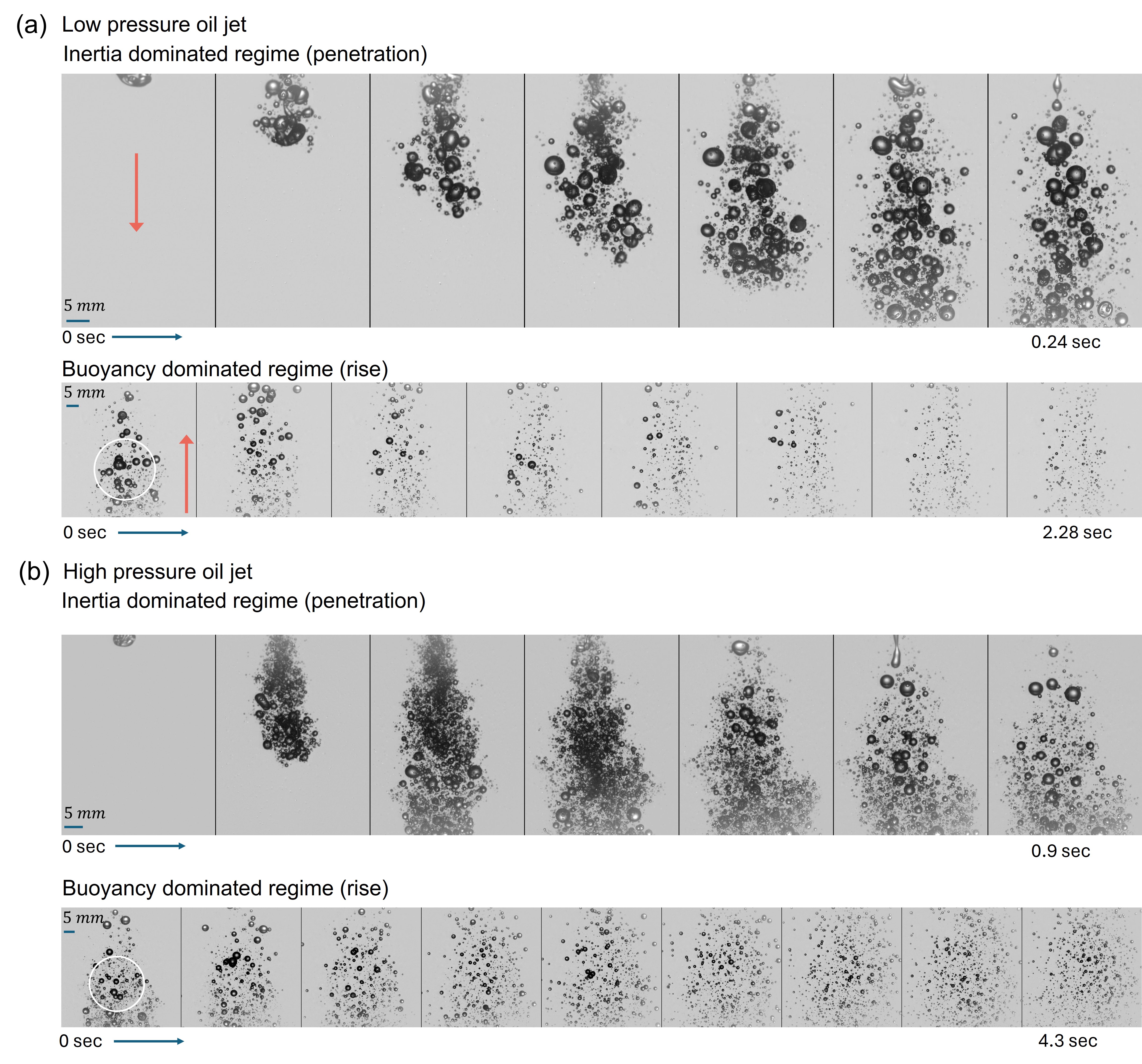}
\caption{Front view of the oil jet (coarse distribution) created through (a) a low compression ratio and (b) a high compression ratio. The first row of both (a) and (b) depicts time-series snapshots as the jet penetrates through the water depth by virtue of the inertia provided during formation. The second row of both (a) and (b) shows the rise of the droplets when the buoyancy force becomes dominant. The white circular region qualitatively depicts the zone that will be covered by a typical vortex ring. The jet created through the higher compression ratio penetrates for a longer time, as can be seen from the time values. The orange arrows indicate the direction of the jet swarm of droplets.}\label{fig:jet}
\end{figure}

Figure~\ref{fig:jet} shows the front view of the oil jet used in the present study. During the inertia-dominated phase, the jet penetrates the water. This phase is longer for high-pressure jets (as can be seen in Fig.~\ref{fig:jet}b) since more inertia is provided. Once the inertia force subsides, the buoyancy force starts to dominate, resulting in the rise of the droplets. The larger droplets can be seen to rise early since they penetrate less and rise faster. Very shortly after this, the droplets achieve terminal velocity, and then the interaction with the vortex ring takes place.

\end{document}